\documentclass[twocolumn,aps,superscriptaddress,showpacs,nofootinbib,floatfix]{revtex4}
\usepackage{epsfig,bm,feynmf}
\usepackage{graphics}
\usepackage{amsmath}
\usepackage[normalem]{ulem}  
\usepackage[dvips]{color} 

\renewcommand\sout{\bgroup \color{red} \ULdepth=-.5ex \ULset}
\begin{document}


\title{Tomography of the Quark-Gluon-Plasma by Charm Quarks }


\author{Taesoo Song}\email{song@fias.uni-frankfurt.de}
\affiliation{Institute for Theoretical Physics, Johann Wolfgang Goethe Universit\"{a}t, Frankfurt am Main, Germany}
\affiliation{Frankfurt Institute for Advanced Studies, Johann Wolfgang Goethe Universit\"{a}t, Frankfurt am Main, Germany}

\author{Hamza Berrehrah}
\affiliation{Institute for Theoretical Physics, Johann Wolfgang Goethe Universit\"{a}t, Frankfurt am Main, Germany}
\affiliation{Frankfurt Institute for Advanced Studies, Johann Wolfgang Goethe Universit\"{a}t, Frankfurt am Main, Germany}

\author{Daniel Cabrera}
\affiliation{Institute for Theoretical Physics, Johann Wolfgang Goethe Universit\"{a}t, Frankfurt am Main, Germany}
\affiliation{Frankfurt Institute for Advanced Studies, Johann Wolfgang Goethe Universit\"{a}t, Frankfurt am Main, Germany}

\author{Juan~M. Torres-Rincon}
\affiliation{Subatech, UMR 6457, IN2P3/CNRS,
  Universit$\acute{e}$ de Nantes, $\acute{E}$cole des Mines de Nantes,
  4 rue Alfred Kastler, 44307 Nantes cedex 3,   France }

\author{Laura Tolos}
\affiliation{Frankfurt Institute for Advanced Studies, Johann Wolfgang Goethe Universit\"{a}t, Frankfurt am Main, Germany}
\affiliation{Institut de Ciencies de l'Espai (IEEC/CSIC), Campus Universitat Autonoma de Barcelona,
Carrer de Can Magrans, s/n, E-08193 Bellaterra, Spain}

\author{Wolfgang Cassing}
\affiliation{Institut f\"{u}r Theoretische Physik, Universit\"{a}t Gie$\beta$en, Germany}

\author{Elena Bratkovskaya}
\affiliation{Institute for Theoretical Physics, Johann Wolfgang Goethe Universit\"{a}t, Frankfurt am Main, Germany}
\affiliation{Frankfurt Institute for Advanced Studies, Johann Wolfgang Goethe Universit\"{a}t, Frankfurt am Main, Germany}


\begin{abstract}
We study charm production in ultra-relativistic heavy-ion collisions by using the Parton-Hadron-String Dynamics (PHSD) transport approach.  The
initial charm quarks are produced by the Pythia event generator tuned to fit the transverse momentum spectrum and rapidity distribution of
charm quarks from Fixed-Order Next-to-Leading Logarithm (FONLL) calculations.
The produced charm quarks  scatter in the quark-gluon plasma (QGP) with the off-shell partons whose masses and widths are given by the
Dynamical Quasi-Particle Model (DQPM), which reproduces the lattice QCD equation-of-state in thermal equilibrium. The relevant cross sections
are calculated in a consistent way by employing the effective propagators and couplings from the DQPM.  Close to the critical energy
density of the phase transition, the charm quarks are hadronized into $D$ mesons through coalescence and fragmentation. The hadronized $D$ mesons then interact with the
various hadrons in the hadronic phase with cross sections calculated in an effective lagrangian approach with heavy-quark spin symmetry.
Finally, the nuclear modification factor $\rm R_{AA}$ and the elliptic flow $v_2$ of $D^0$ mesons from PHSD are compared with the experimental data
from the STAR Collaboration for Au+Au collisions at $\sqrt{s_{\rm NN}}$ =200 GeV.
We find that in the PHSD the energy loss of $D$ mesons at high $p_T$ can be dominantly attributed to partonic scattering while the actual
shape of $\rm R_{AA}$ versus $p_T$ reflects the  heavy-quark hadronization scenario, i.e. coalescence versus fragmentation. Also the hadronic rescattering is important for the $\rm R_{AA}$ at low $p_T$ and enhances the $D$-meson elliptic flow $v_2$.
\end{abstract}


\maketitle

\section{introduction}
According to the fundamental theory of strong interactions, the Quantum Chromo Dynamics (QCD) \cite{lQCD}, matter changes its phase at high
temperature and density and bound (colorless) hadrons dissolve to interacting (colored) quarks and gluons -- Quark-Gluon-Plasma (QGP).
Such extreme conditions have existed in the early expansion of the universe and now can be realized in the laboratory by collisions of
heavy-ions at ultra-relativistic energies. The study of the phase boundary and the properties of the QGP are the main goal of several
present and future heavy-ion experiments at SPS (Super Proton Synchrotron), RHIC (Relativistic Heavy-Ion Collider), LHC (Large Hadron
Collider) and the future FAIR (Facility for Antiproton and Ion Research) and NICA (Nuclotron-based Ion Collider  fAcility)
\cite{QM2014}.  Since the QGP is created only for a short time (of a couple of fm/c) it is quite challenging to study its properties and to
find the most sensible probes. In order to study the full complexity of the underling problem, one needs to obtain comprehensive information by
the measurement of the 'bulk' light hadrons, electromagnetic probes (dileptons and photons), heavy mesons and jets. The advantage of
the 'hard probes' such as the mesons containing  heavy quarks (charm and beauty) is, firstly, that due to the heavy masses they  are
dominantly produced in the very early stages of the reactions with large energy-momentum transfer, contrary to the light hadrons and
electromagnetic probes.  Secondly, they are not in an equilibrium with the surrounding matter due to smaller interaction cross sections
relative to the light quarks and, thus, may provide an information on their creation mechanisms. Moreover, due to the hard scale,
perturbative QCD (pQCD) should be applicable for the calculation of heavy quark production. As shown in Ref. \cite{Vogt}, the FONLL calculations
are in good agreement with the experimental observables on charm meson spectra in p+p collisions.  This
provides a solid reference frame for studying the heavy-meson production in heavy-ion collisions.

The collective properties of open charm mesons have been addressed experimentally by measuring the nuclear modification factor $\rm R_{AA}$,
which is the ratio of the transverse momentum distribution in $A+A$ collisions relative to p+p collisions scaled by the number of binary
collisions, as well as the collective elliptic flow $v_2$. These two observables are comprehensive since the high $p_T$ part of $\rm R_{AA}$ is
very sensitive to the energy loss of charm quarks (or mesons) during their propagation through the partonic (or hadronic) medium due to the
interaction processes. The low $p_T$ part is more sensitive to the hadronization mechanism and, thus, provides constraints on the relative
scale for the transition from partonic to hadronic degrees-of-freedom.
Moreover, the elliptic flow $v_2$ of charm quarks characterizes the collectivity of the system developed from the very early stage to the freeze-out time
thereby including interactions from the partonic and hadronic phase. Thus, the final momentum distribution of the heavy mesons is sensitive
to the details of the expansion of the plasma itself as well as to the strength of the interaction of the heavy quark with the partons and the
formed $D$-mesons with the hadronic environment.  It depends on the density of partons (and their properties) present in the QGP and hence
on whether the plasma is in equilibrium during the expansion.

The first charm measurements at RHIC energies by the PHENIX\cite{eePHENIX} and STAR \cite{eeSTAR} collaborations were
related to the single non-photonic electrons emitted from the decay of charm mesons. However, recently the STAR Collaboration measured
directly the nuclear modification factor and the elliptic flow of $D^0$ mesons in Au+Au collisions at $\sqrt{s_{\rm NN}}=$200
GeV~\cite{Adamczyk:2014uip,Tlusty:2012ix} which allows for a straight forward comparison with the theoretical model calculations. It has been
observed that the $\rm R_{AA}$ and $v_2$ of charm mesons show a similar behavior as in case of light hadrons contrary to expectations from pQCD. Similar observations have been made at LHC energies, too \cite{expLHC}.

It  still remains a challenge for the theory to reproduce the experimental data and to explain simultaneously the large energy loss
of charm quarks ($\rm R_{AA}$) and the strong collectivity ($v_2$) -- cf. e.g. Refs.
\cite{GolamMustafa:1997id,Moore:2004tg,Zhang:2005ni,Molnar:2006ci,
vanHees:2005wb,Gossiaux:2010yx,Gossiaux:2012ya,Ozvenchuk:2014rpa,Cao:2013ita,
Alberico:2011zy,Sharma:2009hn,He:2011qa,Akamatsu:2008ge,BAMPS,
Lang:2012yf,Das:2013kea,Das:2015ana}.
The interactions of charm quarks with the partonic medium are commonly based on pQCD with massless light quarks and a fixed or running coupling.  The time
evolution of the charm-quark distribution in the expanding fireball is approximated by the Fokker-Plank equation where the response of the
partonic (or hadronic) medium is expressed in terms of temperature- and momentum-dependent drag and diffusion coefficients. The modeling of the
temperature profile is often done in a fireball model or by hydrodynamic calculations (ideal or viscous) which start with some
initial conditions and follow the dynamical evolution according to the chosen Equation-of-State (EoS) under the assumption of local
equilibrium. In partonic cascade models one solves the Boltzmann equation for massless quarks/gluons with the pQCD cross sections for
some fixed coupling $\alpha_s$.  The hadronization of charm quarks is done assuming coalescence at low $p_T$ and fragmentation at high
$p_T$.  In spite that many models may describe the $\rm R_{AA}$, it is still difficult to obtain simultaneously a consistent description of
the elliptic flow $v_2$ using the same assumptions and model parameters \cite{Das:2015ana}, e.g. the choice of the running coupling, the
K-factor to scale the pQCD cross sections, the time evolution profile,
etc.  Moreover, the conclusions on the amount of suppression due to collisional energy loss by means of the elastic interactions of charm
quarks with the QGP partons versus the radiative energy loss due to the emission of soft gluons (i.e. gluon bremsstrahlung) are still far from
being robust.  Also the influence of hadronization and especially hadronic rescattering is not yet settled, too. Moreover, the results
turned out to be also sensitive to the time evolution models involved.

Our goal here is to study the charm dynamics based on a consistent microscopic transport approach for the charm production, hadronization
and rescattering with the partonic and hadronic medium. In this study we will confront our calculations within the
Parton-Hadron-String Dynamics (PHSD) approach to the experimental data on charm at RHIC energies and discuss the perspectives/problems of
using the charm quarks for the tomography of the QGP.  To achieve this goal we embed the heavy-quark physics in the existing  PHSD transport
approach \cite{PHSD} which incorporates explicit partonic degrees-of-freedom in terms of strongly interacting quasiparticles
(quarks and gluons) in line with an equation-of-state from lattice QCD (lQCD)
as well as dynamical hadronization and hadronic elastic and inelastic collisions in
the final reaction phase. Since PHSD has been successfully applied to
describe the final distribution of mesons (with light quark content)
from lower SPS up to LHC energies \cite{PHSD,PHSDrhic,Volo,Linnyk}, it
provides a solid framework for the description of the creation,
expansion and hadronization of the QGP as well as the hadronic
expansion with which the heavy quarks interact either as quarks or as
bound states such as $D$-mesons.

Thus, the principle differences of our approach to the previous
dynamical models (including earlier HSD \cite{HSD} studies on the charm
dynamics \cite{review}) are: \\

\noindent (i) the degrees-of-freedom for the QGP are massive and strongly
interacting quasiparticles contrary to massless and weakly interacting
pQCD partons; \\

\noindent (ii) a non-equilibrium off-shell microscopic transport approach is
employed for the QGP dynamics, hadronization  and the hadronic phase  instead of
simplified descriptions of the parton dynamics in terms of Fokker-Planck
equations + QGP hydrodynamics (assuming local equilibration) or
Boltzmann-type partonic cascades with massless light quarks; \\

\noindent (iii) rescattering of $D$-mesons in the hadronic phase
in line with an up-to-date effective Lagrangian approach from Ref. \cite{Tolos:2013kva}.\\

This paper is organized as follows:
In Sec. \ref{PHSD} we recall the basic ideas of the PHSD approach while in
Sec.~\ref{initial} we describe
the production of initial charm quark pairs in hard binary
nucleon-nucleon  collisions and their implementation in PHSD. In
Sec.~\ref{QGP} we present the interactions of charm quarks with
off-shell partons in the QGP calculated earlier in Refs.
\cite{Berrehrah:2013mua,Berrehrah:2014kba,Hamza14}, the hadronization of charm
quarks in Sec.~\ref{tc}, and the $D$ meson interactions with hadrons in
Sec.~\ref{HG2} that are based on the cross sections from Ref.
\cite{Tolos:2013kva}. Finally, the nuclear modification factor $\rm R_{AA}$
and the elliptic flow $v_2$ of $D$ mesons from PHSD are presented in
Sec.~\ref{results} and compared to the available data.  A summary
completes this work in Sec.~\ref{summary}.

\section{The PHSD transport approach}\label{PHSD}

The Parton-Hadron-String Dynamics (PHSD) transport
approach~\cite{PHSD,PHSDrhic} is a microscopic covariant dynamical model
for strongly interacting systems formulated on the basis of
Kadanoff-Baym equations \cite{Kadanoff1,Kadanoff2} for
Green's functions in phase-space
representation (in first order gradient expansion beyond the
quasiparticle approximation). The approach consistently describes
the full evolution of a relativistic heavy-ion collision from the
initial hard scatterings and string formation through the dynamical
deconfinement phase transition to the strongly-interacting
quark-gluon plasma (sQGP) as well as hadronization and the
subsequent interactions in the expanding hadronic phase as in
the Hadron-String-Dynamics (HSD) transport approach \cite{HSD}.
The transport theoretical description of quarks and gluons
in the PHSD is based on the Dynamical Quasi-Particle Model
(DQPM) for partons that is constructed to reproduce lQCD results for a quark-gluon plasma in thermodynamic
equilibrium~\cite{Cassing:2008nn} on the basis of effective propagators for quarks and gluons.
The DQPM is thermodynamically consistent and the effective parton propagators
incorporate finite masses (scalar mean-fields) for
gluons/quarks as well as a finite width that describes the medium dependent reaction rate.
For fixed thermodynamic parameters $(T, \mu_q)$ the partonic width's $\Gamma_i(T,\mu_q)$  fix the effective two-body interactions that are presently implemented in the PHSD~\cite{Vitaly}. The PHSD differs
from conventional Boltzmann approaches in a couple of essential aspects: i) it incorporates dynamical
quasi-particles due to the finite width of the spectral functions (imaginary part of the propagators);
ii) it involves scalar mean-fields that substantially drive the collective flow
in the partonic phase; iii) it is based on a realistic equation of state from lattice QCD and thus
describes the speed of sound $c_s(T)$ reliably; iv) the hadronization is described by the fusion of off-shell
partons to off-shell hadronic states (resonances or strings) and does not violate the second law of thermodynamics;
v) all conservation laws (energy-momentum, flavor currents etc.) are fulfilled in the hadronization
contrary to coalescence models; vi) the effective partonic cross sections no longer are given by pQCD and are 'defined' by the DQPM in a consistent fashion and probed by transport coefficients (correlators) in thermodynamic equilibrium by performing PHSD calculations in a finite box with periodic boundary conditions (shear- and bulk viscosity,
electric conductivity, magnetic susceptibility etc. \cite{Vitaly2,Ca13}).

In the beginning of relativistic heavy-ion collisions color-neutral
strings (described by the FRITIOF LUND model~\cite{FRITIOF})
are produced in hard scatterings of nucleons from
the impinging nuclei. These strings are dissolved into
'pre-hadrons' with a formation time of $\sim$ 0.8 fm/c in their rest
frame, except for the 'leading hadrons', i.e. the fastest residues of the
string ends, which can re-interact (practically instantly) with hadrons with a reduced
cross sections in line with quark counting rules.
If, however, the local energy density is larger than the
critical value for the phase transition, which is taken to be $\sim$
0.5 ${\rm GeV/ fm^3}$, the pre-hadrons melt into (colored) effective quarks
and antiquarks in their self generated repulsive
mean-field as defined by the DQPM~\cite{Cassing:2008nn}.
In the DQPM the quarks, antiquarks and gluons are dressed quasiparticles
and have temperature-dependent effective masses and widths
which have been fitted to lattice thermal
quantities such as energy density, pressure and entropy density.
The nonzero width of the quasiparticles implies the off-shellness of partons, which is taken
into account in the scattering and propagation of partons in the QGP on
the same footing (i.e. propagators and couplings).
We point out that the DQPM does not
provide effective propagators for the $c-$ and ${\bar c}-$-quarks since the latter degrees-of-freedom
are subdominant in the entropy density for temperatures of 1 - 3 $T_c$ due to their large mass
and thus cannot be determined properly by the entropy from the lattice. Nevertheless, the interactions
of the charm quarks with the light ($u,d,s)$ quarks can be computed on the basis of the running coupling $g^2(T/T_c)$ in the DQPM and the effective propagotors for the light quarks. As demonstrated in Ref. \cite{Berrehrah:2013mua} the resulting
differential cross sections of $c, {\bar c}$ quarks only very slightly depend on the spectral width of the charm
quarks such that even the on-shell limit for these degrees-of-freedom with a mass
 of $\sim$ 1.5 GeV provides a very reasonable approximation \cite{Berrehrah:2013mua,Berrehrah:2014kba}.

The transition from the partonic to hadronic degrees-of-freedom (for light quarks/antiquarks)
is described by covariant transition rates
for the fusion of quark-antiquark pairs to mesonic resonances or three quarks (antiquarks) to baryonic states,
i.e. by the dynamical hadronization. We already mention here that this hadronization process is restricted to 'bulk'
transverse momenta $p_T$  up to $\sim $ 2 GeV and has to be replaced by fragmentation for high $p_T$. Note that due
to the off-shell nature of both partons and hadrons, the
hadronization process described above obeys all conservation laws (i.e.
four-momentum conservation and flavor current conservation) in each event, the detailed balance relations
and the increase in the total entropy $S$. In the hadronic phase PHSD is equivalent to the
hadron-strings dynamics (HSD) model \cite{HSD} that has been employed in the past from SchwerIonen-Synchrotron (SIS)
to SPS energies. On the other hand the PHSD approach has been applied to p+p, p+A and relativistic heavy-ion collisions from lower SPS to LHC
energies and been successful in describing a large number of
experimental data including single-particle spectra, collective flow as well as electromagnetic probes~\cite{PHSD,PHSDrhic,Volo,Linnyk}.

In this study we use the Pythia event generator~\cite{Sjostrand:2006za} for nucleon-nucleon binary collisions
to produce charm-quark pairs in relativistic heavy-ion collisions and
evolve the dynamics of the charm quarks within PHSD in order to
understand the dominant mechanisms in comparison to the recent
experimental data on $D^0$ mesons from STAR. Whereas the dynamics of the charm quarks
is essentially Boltzmann-like, i.e. without mean-fields and dynamical width, their interactions
with the dynamical partons (light quarks and gluons) is based on the effective coupling from the DQPM and the DQPM propagators \cite{Berrehrah:2013mua,Berrehrah:2014kba}.

\section{Initial charm quark production}\label{initial}

Before studying the charm production in relativistic heavy-ion
collisions, we discuss the  charm production in p+p collisions at the
top RHIC energy of $\sqrt{s_{\rm NN}}$ = 200 GeV.  The charm production in
p+p collisions also plays the role as a reference for the nuclear
modification factor $\rm R_{AA}$ in heavy-ion collisions.

\begin{figure}[h]
\centerline{
\includegraphics[width=9.5 cm]{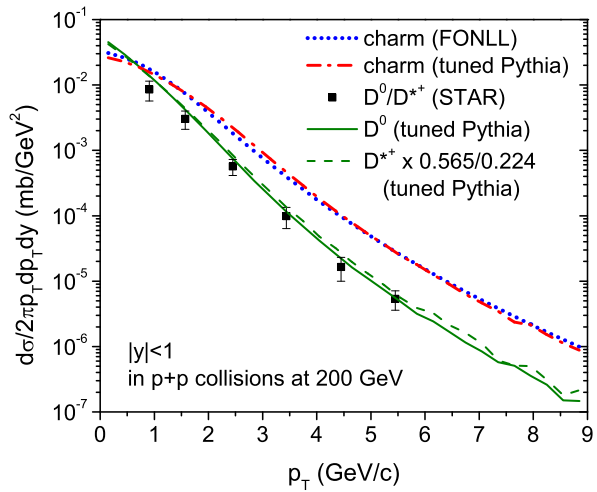}}
\centerline{
\includegraphics[width=9.5 cm]{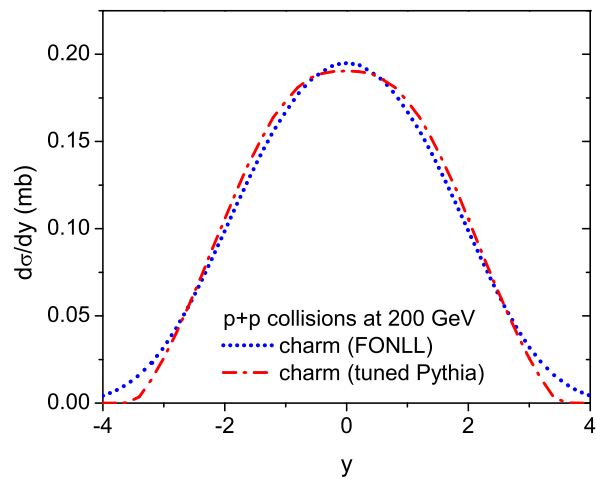}}
\caption{(Color online) Transverse momentum (upper) and rapidity (lower)
distributions of charm quarks in p+p collisions at $\sqrt{s_{\rm NN}}=$200 GeV from FONLL (dotted lines) and the tuned Pythia event generator (dot-dashed lines); in the upper part the transverse momentum spectrum of $D^0$ mesons which are
fragmented from charm quarks with the contribution from $D^*$ decay included (solid line) and that of $D^{*+}$ after scaling are compared with the experimental data from the STAR Collaboration~\cite{Adamczyk:2012af}.}
\label{pp1}
\end{figure}

We use the Pythia event generator to produce charm and anticharm quarks
in p+p collisions with the parameters PARP(91)=1.0 GeV/c and
PARP(67)=1.0 as in Ref.~\cite{Sjostrand:2006za,Adamczyk:2012af}. The
former parameter is the Gaussian width of the primordial transverse
momentum of a parton which initiates a shower in
hadrons and the latter the parton shower level parameter. We note that
by an additional suppression of the transverse momenta of charm and
anticharm quarks by 10 \% and a suppression of rapidities by 16 \%  the
transverse momentum spectrum as well as rapidity distribution of charm
and anticharm quarks from the Pythia event generator are very similar
to those from the FONLL calculations in p+p collisions at $\sqrt{s_{\rm
NN}}=$200 GeV as shown in Fig.~\ref{pp1}. Here the
red dot-dashed lines are from
the Pythia event generator (after tuning) and the blue dotted lines from the
FONLL calculations, respectively.

The produced charm and anticharm quarks  hadronize by emitting soft
gluons.  The probabilities for a charm quark to  hadronize into
$D^0,~D^+,~D^{*0},~D^{*+},~D_s^+$, and $\Lambda_c$ are, respectively, taken to be 0.2, 0.174, 0.213, 0.224, 0.08, and 0.094 ~\cite{Amsler:2008zzb}.  The momentum of the hadronized $D$ meson
or $\Lambda_c$ is given by the fragmentation function~\cite{Peterson:1982ak},
\begin{eqnarray}
D_Q^H(z)\sim \frac{1}{z[1-1/z-\epsilon_Q/(1-z)]^2},
\end{eqnarray}
where $z$ is the momentum fraction of the hadron $H$ fragmented from
the heavy quark $Q$ while $\epsilon_Q$ is a fitting parameter which is
taken to be $\epsilon_Q$ = 0.01 in our study. We note that the
parameter $\epsilon_Q$ used here is smaller than the usual value
because the transverse momentum of charm quarks is reduced in our
study. The solid line in Fig.~\ref{pp1} (upper part) shows the transverse momentum
spectrum of $D^0$ mesons after charm quark fragmentation including the
contribution from the decay of $D^{*0}$ and $D^{*+}$. We can see that
our results reproduce the experimental data from the STAR
Collaboration  \cite{Adamczyk:2012af} reasonably well.

\begin{figure}[h]
\centerline{
\includegraphics[width=9.5 cm]{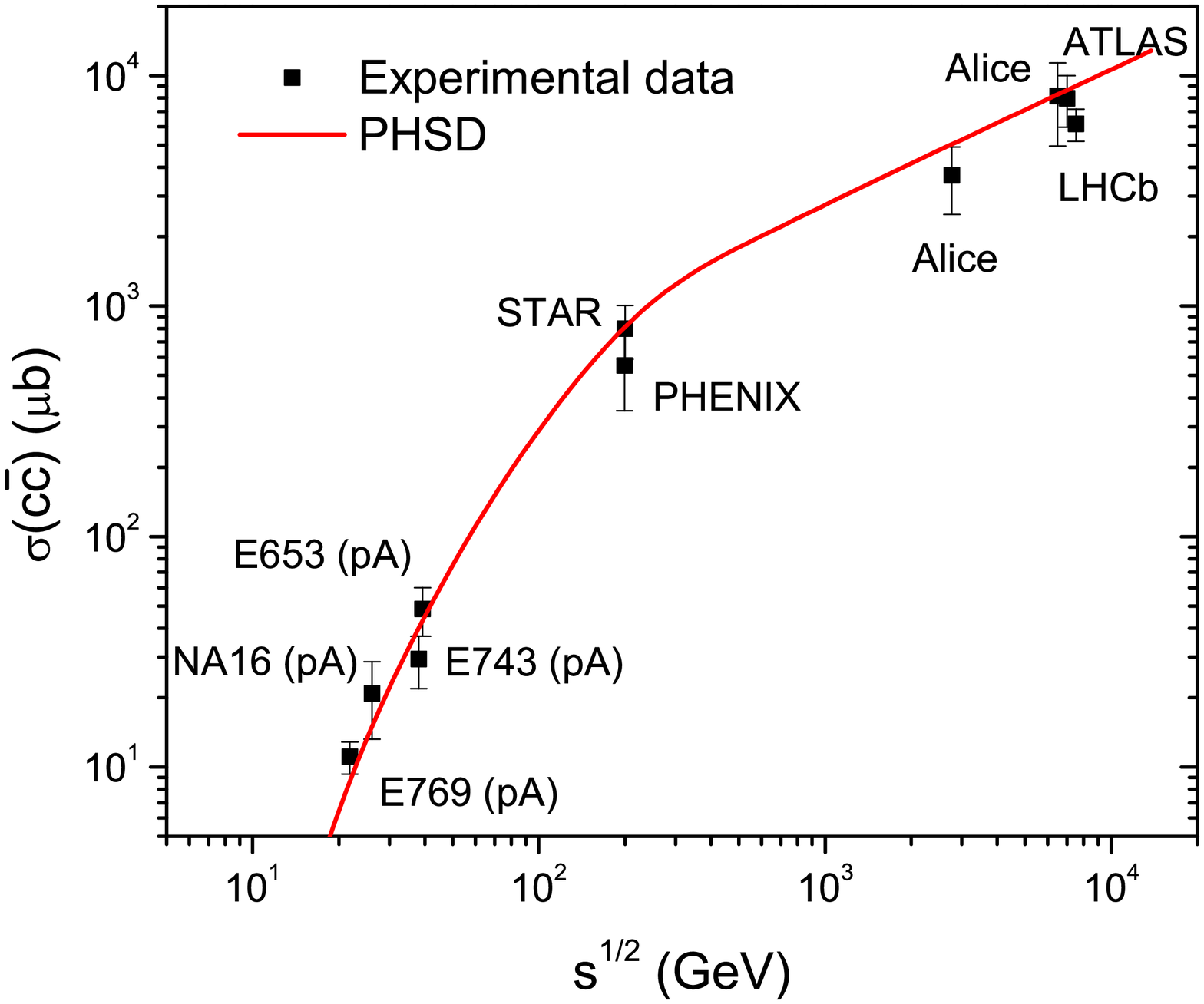}}
\centerline{
\includegraphics[width=9.5 cm]{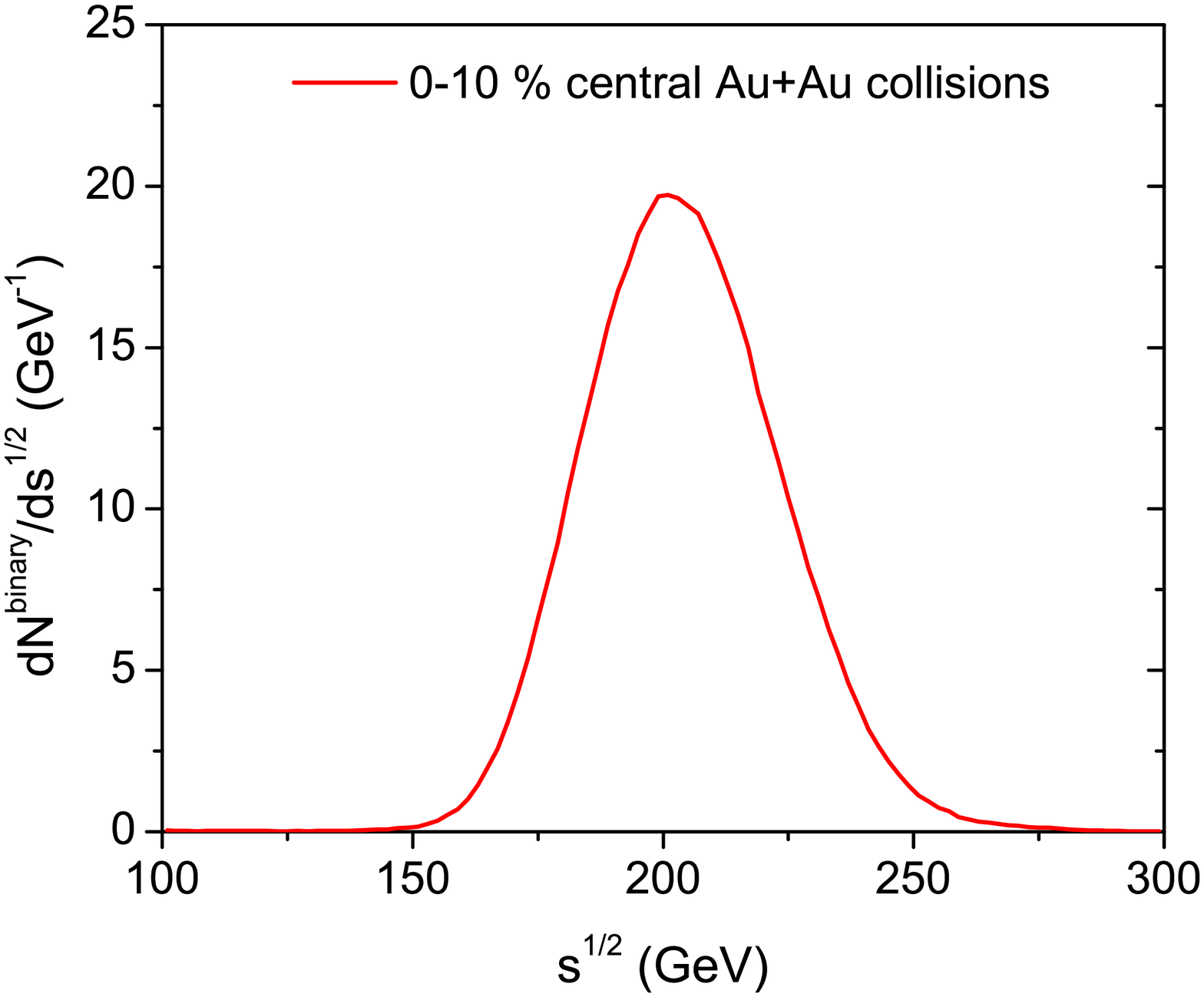}}
\caption{(Color online) The total cross section for charm production in
p+p collisions (as parameterized in the PHSD) is compared with the
experimental data at various collision
energies~\cite{Adamczyk:2012af,delValle:2011ex} (upper panel); the
distribution of binary nucleon-nucleon collision energies in 0-10 \%
central Au+Au collisions at $\sqrt{s_{\rm NN}}=$200 GeV from PHSD
including Fermi smearing (lower panel).}
\label{pp2}
\end{figure}

In heavy-ion collisions, the nucleons in the target and projectile
nuclei are distributed in coordinate space according to the nuclear density
distribution given by
\begin{eqnarray}
\rho(r)\sim\frac{1}{1+\exp[(r-r_0)/a]},
\end{eqnarray}
where $r_0=1.124 A^{1/3}$ and $a=0.02444 A^{1/3}+0.2864$ with $A$
being the mass number of the target or projectile nucleus.
Each nucleon has Fermi motion depending on the local nucleon density
which is chosen randomly by Monte Carlo.  Although the Fermi momentum
is small in the rest frame of each nucleus, the component along the
beam direction becomes large in the laboratory frame due to the Lorentz
transformation ($\gamma_{cm} \approx$ 100 at the top RHIC energy).
Accordingly, the Fermi motions smears the energy of nucleon-nucleon
binary collisions in relativistic heavy-ion collisions. The lower part
of Fig.~\ref{pp2} shows the distribution of binary nucleon-nucleon
collision energies in 0-10 \% central Au+Au collisions at $\sqrt{s_{\rm NN}}=$200 GeV from PHSD. We mention
that this Fermi smearing is usually neglected in theoretical models for
charm production and propagation which is not crucial at RHIC energies but becomes important
close to threshold energies for charm production.

Since charm-quark production requires a high energy-momentum transfer,
the number of produced charm quark pairs in relativistic heavy-ion
collisions is proportional to the number of binary nucleon-nucleon
collisions $N_{bin}$. Since the probability to produce a charm quark
pair depends on invariant energy and is less than that for primary hard
collision in the PHSD, the binary nucleon-nucleon collisions producing
charm quark pairs are chosen by Monte Carlo from the ratio
of the cross section for charm production in nucleon-nucleon collision, $\sigma_{c\bar{c}}^{pp}(\sqrt{s})$, to the inelastic nucleon-nucleon cross section. The total cross
section in PHSD for charm production in nucleon-nucleon collisions is
parameterized by
\begin{eqnarray}
\sigma_{c\bar{c}}^{pp}(\sqrt{s})=A[1-\sqrt{(s_0/s})]^\alpha(s/s_0)^\beta,
\end{eqnarray}
where $A$, $\alpha$ and $\beta$ are fit parameters.
We use two different parameter sets, one for  $\sqrt{s}\le $200 GeV and the other for $\sqrt{s} > $200 GeV  to fit the experimental data covering a wide range of collision energies as shown in the upper part of Fig.~\ref{pp2}; both parameterizations are smoothly connected at $\sqrt{s}= $200 GeV.

The energy-momenta of the produced charm and anticharm quarks in each
collision event are given by the Pythia event generator as in p+p
collisions but for the actual (smeared) collision energy. In the
Pythia event generator, the charm and anticharm quarks are produced in
the center-of-mass frame of two colliding nucleons with the incident
nucleons being aligned along the $z$-direction. Therefore in PHSD  we
rotate the momenta of the generated charm and anticharm quark to the
original orientation of the two incident nucleons in their
center-of-mass frame and then boost to the calculational frame. We assume
that the two nucleons, which produce the charm quark pair, keep their
transverse momentum and loose only longitudinal momentum such that the
total energy (including the produced charm quark pair) is conserved and
the two nucleons are still on-shell after the hard collision. Although
this prescription seems to violate spatial momentum conservation,
we have to mention that this violation is very small since the charm quark pair is produced
together with several (plenty) light hadrons in the same event which balance the spatial momentum.

\section{Charm-quark scattering in the QGP}\label{QGP}

\subsection{$q+c$ and $g+c$ elastic scattering cross sections}

Quarks, antiquarks, and gluons are dressed in the QGP and have temperature-dependent effective masses and widths. In the DQPM, the mass and width of the light partons are given by thermal quantum-field theory assuming leading order diagrams but the strong coupling $g^2(T)$ is fitted to lattice data on energy and entropy densities, etc.~\cite{Cassing:2008nn}. Note that a nonzero width of a parton reflects the off-shell nature as well as the strong interaction of the quasi-particle and finite life-time.

The charm and anticharm quarks produced in early hard collisions interact with the dressed off-shell partons in the QGP. The cross sections for the charm quark scattering with massive off-shell partons are calculated considering the mass spectra of final state particles~\cite{Berrehrah:2013mua,Berrehrah:2014kba}. In the current study the charm quark mass is taken to be 1.5 GeV and its mass spectrum is neglected for simplicity. Since the charm quark is heavy, the off-shell effect is small contrary to the light quarks and gluons (see Refs. \cite{Berrehrah:2013mua,Berrehrah:2014kba} for a quantitative study of this effect). Moreover, the off-shell effects from parton spectral functions are small, except for reducing the kinematic threshold.

In the current study we consider only elastic scattering of charm quarks by light quarks and gluons. We do not consider yet the radiative processes which generate radiative energy loss because we expect that, due to the large gluon mass in the DQPM, the radiative processes are sub-dominant as compared to the collisional ones, especially for low charm-quark momenta $(p_T)$. We expect the radiative energy loss to contribute at very high $p_T$ as accessible experimentally at the LHC~\cite{Younus:2013rja}.

\begin{figure}[h]
\centerline{
\includegraphics[width=3.45 cm, height=3.25 cm]{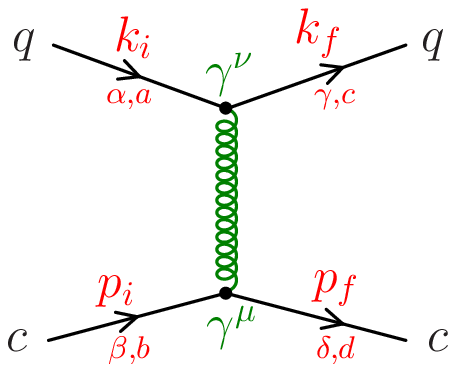}}
\centerline{
\includegraphics[width=8.6 cm, height=3.3 cm]{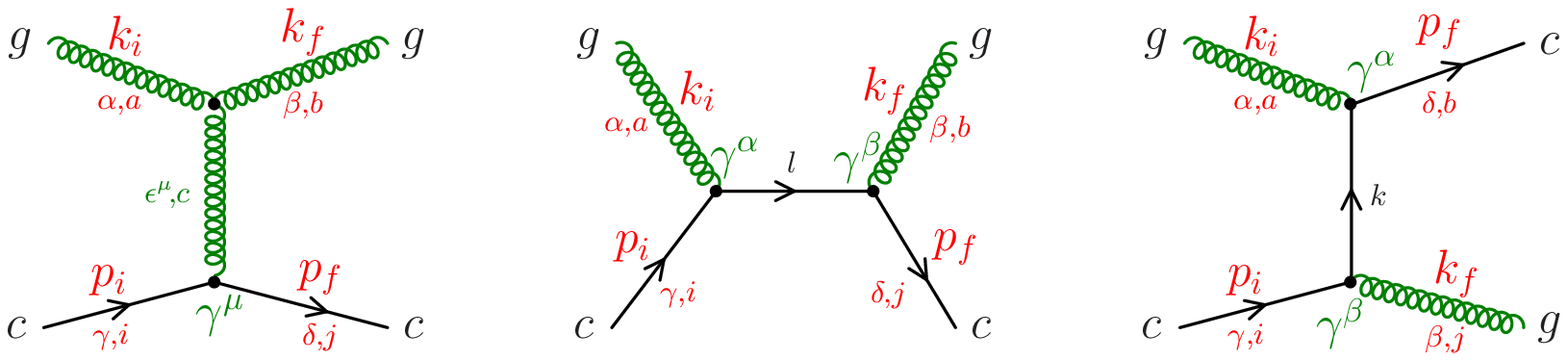}}
\caption{(Color online) Feynman diagrams for the scattering of $q+c \rightarrow q+c$ and $g+c \rightarrow g+c$. Latin (Greek) subscripts denote colour (spin) indices. $k_i$ and $p_i$ ($k_f$ and $p_f$) denote the initial (final) 4-momentum of the light quark or the gluon and the heavy quark, respectively.}
\label{qgQFeynman}
\end{figure}

The elastic scattering of charm quarks in the QGP is treated in our study by including the non-perturbative effects of the strongly interacting quark-gluon plasma (sQGP) constituents, i.e. the large coupling, the multiple scattering etc. The multiple strong interactions of quarks and gluons in the sQGP are encoded in their effective propagators with broad spectral functions. As pointed out above the effective propagators, which can be interpreted as resummed propagators in a hot and dense QCD environment, have been extracted from lattice data in the scope of the DQPM \cite{Cassing:2008nn}. Furthermore, in Refs. \cite{Berrehrah:2013mua,Berrehrah:2014kba} we have evaluated, to the lowest order in the perturbation expansion, the transition amplitudes of the processes $q+c \rightarrow q+c$ and $g+c \rightarrow g+c$ as shown in Fig.~\ref{qgQFeynman}. Contrary to the case of massless gluons where the ``Transverse gauge'' is used, we use the ``Lorentz covariance'' for the case of massive gluons since a finite mass in the gluon propagator allows to fix the 0'th components of the gluon fields $A_a^0 (a=1; \dots ; 8)$ by the spatial degrees of freedom $A_a^k (k =1;2;3)$. Furthermore, the divergence encountered in the $t$-channel – when calculating the total cross sections $(\sigma^{\textrm{qc}})$ and $(\sigma^{\textrm{gc}})$ – is cured self-consistently in our computation since the infrared regulator is given by the finite DQPM gluon mass (and width).

For the scattering of charm quark by the light quark and gluon with finite masses and widths, we take into account the spectral functions of the light quark and gluon, and the temperature-dependent running coupling, $g^2(T)$. The propagators of massive vector gluons with finite lifetime and of charm quark with zero life time are, respectively, given by

\begin{eqnarray}
\label{equ:EqPropag}
G^{\mu \nu} (q) &=& - i \frac{g^{\mu \nu} - q^{\mu} q^{\nu}/m_g^2}{q^2 - m_g^2 + i 2 \gamma_g q_0},\nonumber\\
S (p) &=& \frac{p \!\! / + m_c}{p^2 - m_c^2},
\end{eqnarray}
where $m_g$, $\gamma_g$ are ,respectively, gluon mass and width at finite temperature, and the charm quark mass $m_c$ is taken to be 1.5 GeV.

\begin{figure}[h!]
\centerline{
\includegraphics[width=8.0 cm, height=7 cm]{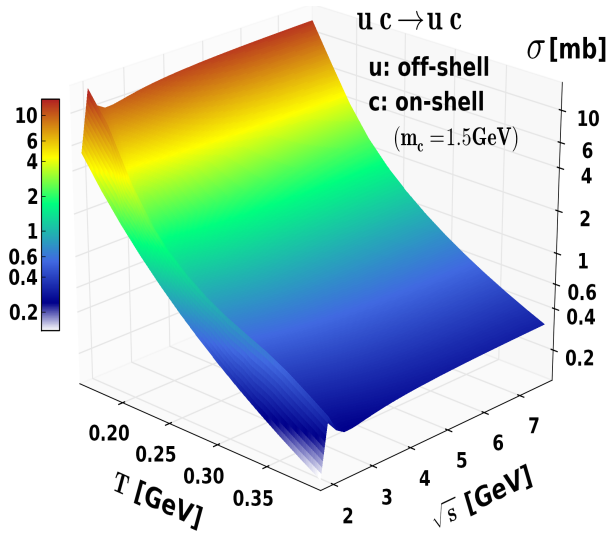}}
\caption{(Color online) Elastic cross section
for $u+c \rightarrow u+c$ scattering as a
function of the temperature $T$ and the invariant energy
$\sqrt{s}$ where the $u$-quark is off-shell and the
charm quark has a constant mass of 1.5 GeV
\cite{Berrehrah:2013mua,Berrehrah:2014kba}.} \label{SigmacuOffTs}
\end{figure}

Fig.~\ref{SigmacuOffTs} shows the elastic cross section of an on-shell charm quark with off-shell $u$-quarks $(\sigma^{\textrm{uc}})$ as a function of the temperature and $\sqrt{s}$, the energy available in the center-of-mass frame. Apart from the threshold region the cross section is rather independent of $\sqrt{s}$, and decreases with increasing temperature mainly due to a smaller coupling strength at high temperature. The large cross section near the critical temperature $T_c = 0.158$ GeV is related to the infrared enhancement of the coupling $\alpha_s (T)$ in the DQPM. The cross section for charm quark and gluon elastic scattering is about twice (9/4 which is the ratio of the different color Casimir operators squared) that for charm quark and light quark scattering.

\begin{figure}[h!]
\centerline{
\includegraphics[width=8.0 cm]{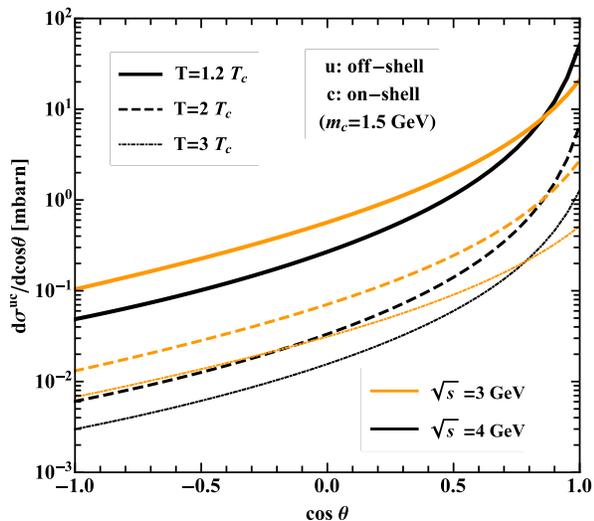}}
\caption{(Color online) Differential elastic cross section for
the scattering of an on-shell
charm quark and off-shell $u$-quarks at $\sqrt{s} = 3$ GeV (orange
lines) and 4 GeV (black lines) for temperatures of 1.2 $T_c$, 2
$T_c$ and 3 $T_c$, with $T_c = 0.158$ GeV. We take 1.5 GeV for the
charm quark mass and DQPM spectral functions (propagators) for the
light  off-shell partonic degrees-of-freedom
\cite{Berrehrah:2013mua}.} \label{dSigmacu}
\end{figure}

Our cross sections also depend on the scattering angle besides collision energy $\sqrt{s}$ and temperature $T$. Fig.~\ref{dSigmacu} presents the differential cross sections $d \sigma/d \cos \theta$ for on-shell charm quark scattering with off-shell $u$-quarks at $\sqrt{s} = 3$ GeV (orange lines) and 4 GeV (black lines) for temperatures of 1.2 $T_c$, 2 $T_c$ and 3 $T_c$. Different from the pQCD-inspired models, $d \sigma/d \cos \theta$ in our 'non-perturbative' model is not so much forward peaked (large enhancement for small angles or small momentum transfers $t$) \cite{Berrehrah:2013mua}. Therefore, efficient momentum transfers more often occur in our partonic scattering as compared to the usual pQCD $2 \rightarrow 2$ scattering \cite{Berrehrah:2014kba}.

\subsection{Charm spatial diffusion coefficient}

Using the transition amplitudes for the elastic scattering of charm
quarks by the partons in medium,
one can calculate the charm spatial diffusion
coefficient $D_s$ from the drag coefficient, $A = \frac{d<\textbf{p}_c>}{d t}$ through $\eta_D =
A/p_c$:
\begin{eqnarray}
D_s =\lim_{p_c\to 0} \frac{T}{m_c \eta_D},
\end{eqnarray}
or from the diffusion coefficient $\kappa = \frac{1}{3} \frac{d<(\textbf{p}_c - \textbf{p}'_c)^2>}{d t}$:
\begin{eqnarray}
\label{equ:Ds}
\hspace{0.5cm} D_s = \lim_{p_c\to 0} \frac{\kappa}{2 m_c^2 \eta_D^2}.
\end{eqnarray}
Both definitions agree with each other if the Einstein relation is valid \cite{Berrehrah:2014kba,Hamza14}.

\begin{figure}[h!]
\begin{center}
\includegraphics[width=7.7 cm]{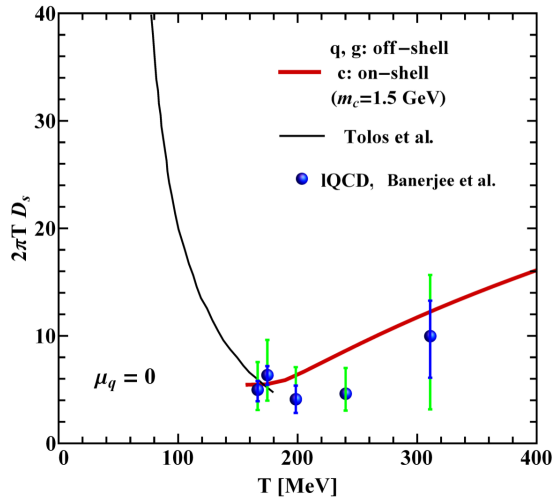}
\caption{(Color online) Spatial diffusion coefficient
of heavy quark, $D_s$, as a function of $T$
for $\mu_q=0$. The black solid line
below $T=180$ MeV is the hadronic diffusion coefficient~\cite{Tolos:2013kva}, and the red solid line above $T_c
\approx 160$ MeV our $D_s$
computation in the partonic environment. The lattice
calculations are from Ref.~\cite{Banerjee:2011ra}.} \label{fig:cDs}
\end{center}
\end{figure}

Using Eq.~(\ref{equ:Ds}), we show in Fig.\ref{fig:cDs}
the spatial diffusion coefficient $D_s$ as a function of $T$
for $\mu_q = 0$. Our results are compared to the
lattice calculations, which have recently been
confirmed by the Bielefeld collaboration~\cite{Banerjee:2011ra},
for temperatures above $T_c\approx$ 160 MeV and with
the spatial diffusion coefficient of a heavy meson in hadronic
matter~\cite{Tolos:2013kva} for temperatures below 180 MeV. We observe that the spatial diffusion coefficients in hadronic
and partonic matters meet each
other and have a pronounced minimum around $T_c$.
Fig.~\ref{fig:cDs} shows that our results agree with those from
the lattice QCD. The smooth transition of the
heavy-quark transport coefficients from the hadronic to the partonic
medium corresponds to a crossover in line with lattice calculations,
and differs substantially from perturbative QCD (pQCD) calculations
which show a large discontinuity  of $D_s$ close to $T_c$
\cite{Berrehrah:2014kba,Hamza14}.

We emphasize that the transport coefficient $D_s$ extracted from our microscopic calculations and its agreement with the lQCD results (within errors) and the corresponding $D$ meson $D_s$ in hadronic medium validate our description for the coupling of charm with the QGP matter.

\begin{figure}[h]
\centerline{
\includegraphics[width=9.5 cm]{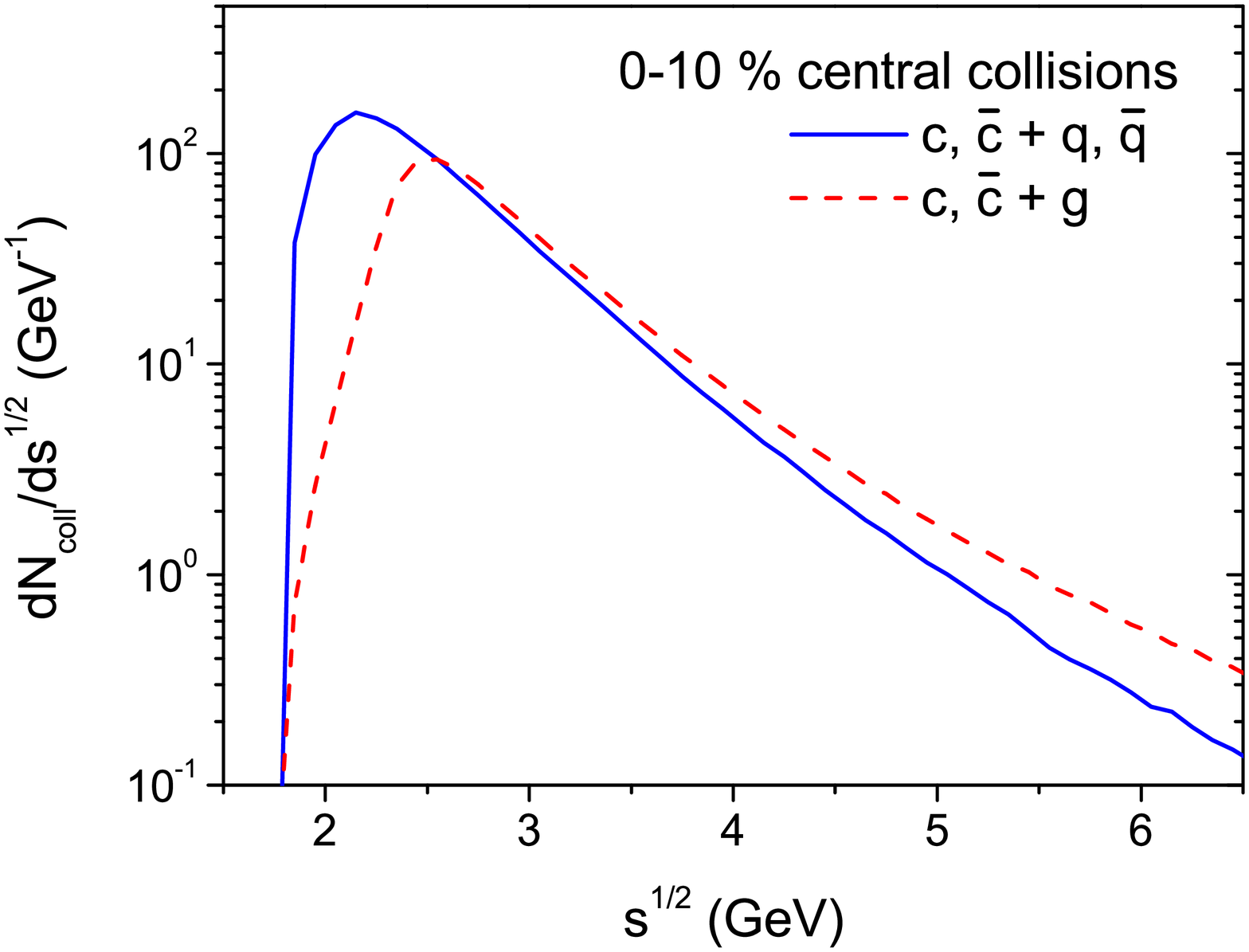}}
\caption{(Color online) The distributions of charm and anticharm quark
scattering by quarks/antiquarks (solid line) and gluons (dashed line)
 as a function of the invariant energy $\sqrt{s}$
in 0-10 \% central Au+Au collisions.}
\label{scatterP}
\end{figure}

Fig.~\ref{scatterP} shows the distributions of charm and anticharm quark
scattering  with quarks/antiquarks (solid line) and gluons (dashed line)
 as a function of the invariant energy $\sqrt{s}$ in 0-10 \% central Au+Au collisions.
The total number of charm and anticharm quark scatterings with light
quarks
and light antiquarks is about 135 and that with gluons is about 83.
We recall that about 19 pairs of charm and anticharm quarks are produced
in this centrality range; each charm or anticharm quark thus experiences
on average 6 elastic scatterings with partons before it is hadronized.
Since the dressed quark/antiquark mass is smaller than the dressed gluon mass,
the peak of the distribution for charm and quark scattering is located at a
lower energy compared to that of charm and gluon scattering.
We additionally note that the number of charm quark scatterings with gluons
is not small compared to that with quarks although the number of gluons is
significantly smaller than that of quarks due to their larger mass.
This is because the cross section for the charm and gluon elastic scattering is about
twice that for charm and light quark elastic scattering.
One characteristic of the partonic scattering of charm quarks is that the
scattering distribution in $\sqrt{s}$ has a long tail up to high
invariant energies beyond that expected in a thermal equilibrium.
This is attributed to the cross section for charm and parton elastic scattering
which does not decrease with increasing $\sqrt{s}$ as shown in Fig.~\ref{SigmacuOffTs}.
Therefore, the partonic scattering is effective for the energy loss of
charm and anticharm quarks at high transverse momentum.

\begin{figure}[h!]
\centerline{
\includegraphics[width=9.5 cm]{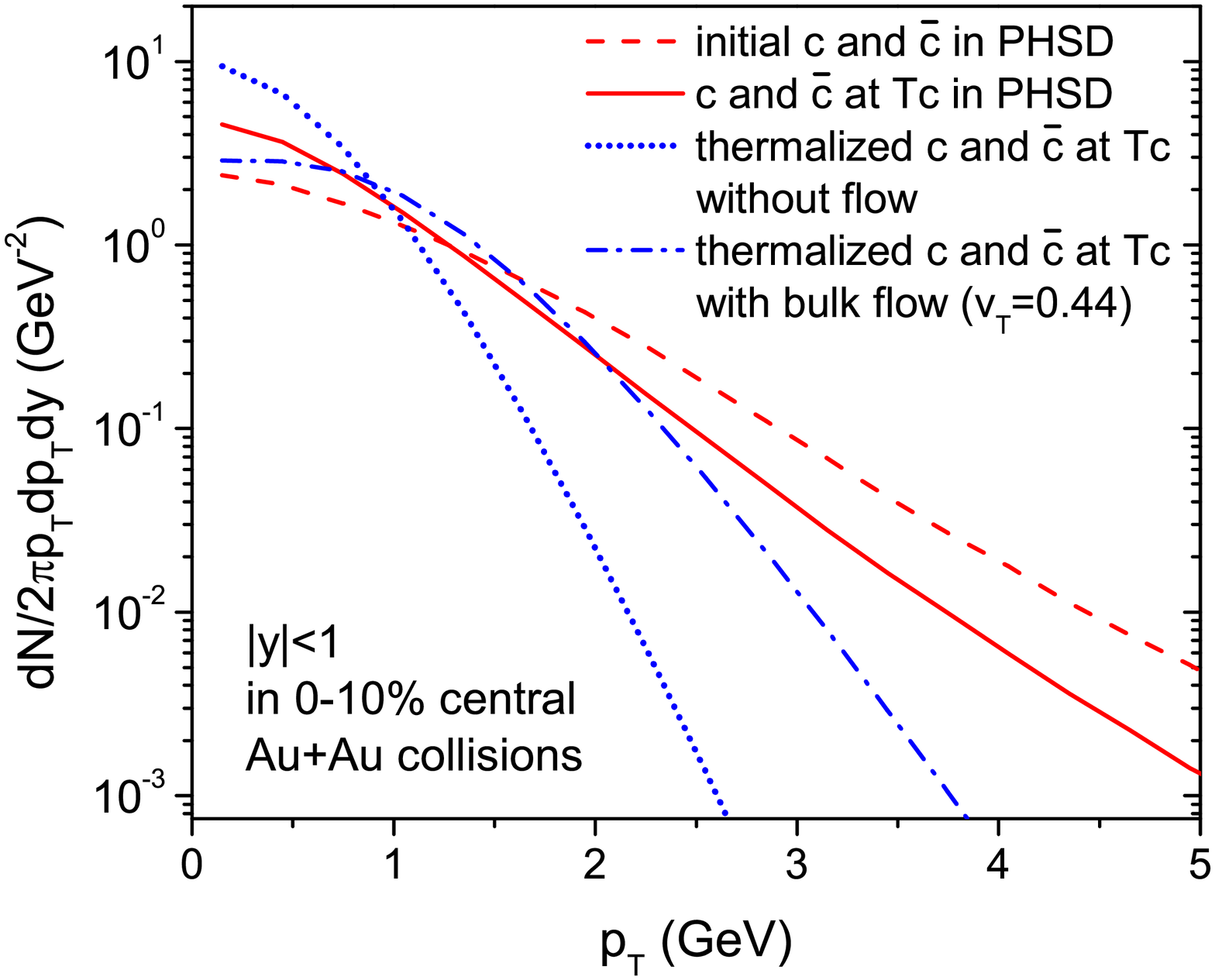}}
\caption{(Color online) Transverse momentum spectra of charm and
anticharm quarks at the initial production (dashed red line) and
at hadronization (solid red line) in 0-10 \% central Au+Au
collisions, compared with those of thermal charm quarks with
(dot-dashed blue line) and without (dotted blue line) considering
the transverse flow of the bulk particles at T=158 MeV which is
the critical temperature for the QGP phase transition in PHSD.}
\label{pt-QGP}
\end{figure}
Fig.~\ref{pt-QGP} compares the transverse-momentum spectrum of
charm and anticharm quarks at the initial production with that at
hadronization in the midrapdity interval $(|y|<1)$. We can see that
the scattering with massive partons softens the transverse
momentum spectrum of charm quarks substantially. These spectra are
also compared in Fig.~\ref{pt-QGP} with the thermal spectrum of
charm and anticharm quarks at $T_c$ with the number of charm and
anticharm quarks being the same as in the PHSD. The
dot-dashed blue line in Fig.~\ref{pt-QGP} shows the spectrum when
we take into account the transverse flow velocity $v_T$ of the bulk
particles which is about 0.44 at $T_c$ in 0-10 \% central
collisions in the PHSD simulations:
\begin{eqnarray}
\frac{dN}{2\pi p_T dp_T dy}\sim m_T I_0(\alpha) K_1(\beta),
\end{eqnarray}
where $m_T=\sqrt{m_c^2+p_T^2}$, $\alpha=p_T\sinh\rho/T$ and $\beta=m_T \cosh\rho/T$ with $\rho=\tanh^{-1}v_T$, and $K_1$ and $I_0$ are modified Bessel functions~\cite{Heinz:2004qz}.

The dotted blue line shows the
spectrum without including the transverse flow ($v_T=0$). Since the
transverse flow velocity of charm quarks in relativistic heavy-ion
collisions is smaller than that of the bulk particles, even though
they are completely thermalized~\cite{Song:2011kw}, the
transverse-momentum spectrum of thermalized charm quarks is
between the dotted blue and the dot-dashed blue lines. This figure
suggests that charm and anticharm quarks are close to thermal
equilibrium at low transverse momentum ($p_T <$ 2 GeV) after
partonic scattering while they are still off-equilibrium at higher
transverse momenta since the solid red and the dot-dashed blue line
start to deviate substantially for higher $p_T$.

\section{Hadronization of charm quarks}\label{tc}

Since the hot and dense  matter created by a relativistic heavy-ion
collision expands with time, the energy density of the matter
decreases and the deconfined degrees-of-freedom hadronize to color
neutral hadronic states. Once the local energy density in PHSD
becomes lower than 0.5 ${\rm GeV/fm^3}$, the partons are
hadronized. First we look for all combinations of a charm quark
and light antiquarks or of an anticharm quark and a light quark and
calculate the probability for each combination to form a $D$ or
$D^*$ (or $D_s, D^*_s$) meson. The probability for a quark
and an anti-quark to form a meson is given by
\begin{eqnarray}
f(\boldsymbol\rho,{\bf k}_\rho)=\frac{8g_M}{6^2}
\exp\left[-\frac{\boldsymbol\rho^2}{\delta^2}-{\bf k}_\rho^2\delta^2\right],
\label{meson}
\end{eqnarray}
where $g_M$ is the degeneracy of the meson $M$, and
\begin{eqnarray}
\boldsymbol\rho=\frac{1}{\sqrt{2}}({\bf r}_1-{\bf r}_2),\quad{\bf k}_\rho=\sqrt{2}~\frac{m_2{\bf k}_1-m_1{\bf k}_2}{m_1+m_2},
\label{coalescence}
\end{eqnarray}
with $m_i$, ${\bf r}_i$ and ${\bf k}_i$ being the mass, position and momentum of the quark or antiquark $i$, respectively. The width parameter $\delta$ is related to the root-mean-square radius of the meson produced through
\begin{eqnarray}
\langle r^2 \rangle=\frac{3}{2}\frac{m_1^2+m_2^2}{(m_1+m_2)^2}\delta^2
\end{eqnarray}
and thus determined by experiment (if available).

The degeneracy factors, $g_M$, for $D$ and $D^*$ mesons are 1 and 3,
respectively. We also
take into account higher excited states of D mesons from the
Particle Data Group~\cite{Agashe:2014kda}: $D_0^*(2400)^0$,
$D_1(2420)^0$, and $D_2^*(2460)^{0,\pm}$. They dominantly decay into
$D+\pi$ or $D^*+\pi$ and each branching ratio is not known yet.
Therefore, we assume that the higher excited state decays immediately after hadronization into
$D+\pi$ or $D^*+\pi$ with the branching ratio of 1 to 3, keeping its
flavor. We further simplify that higher excited states have the same
mass, 2460 MeV, since $D_2^*(2460)^{0,\pm}$ are most abundantly
produced in the coalescence model due to its large spin, and they
are assumed to have the same radius as $D$ and $D^*$ mesons.

In the actual simulations of the PHSD we perform several tens of
events in parallel to obtain smooth local energy and particle number
densities as a function of time; the charm or anticharm quark in the
hadronization is allowed to take its partner from other events.
Therefore, the coalescence probability given in Eq.~(\ref{meson}) is
divided by the number of parallel events such that the results
become independent on the number of parallel runs.

Collecting all possible combinations of a charm or an anticharm quark
with light quarks or antiquarks and calculating the coalescence
probability for each combination from Eq.~(\ref{meson}), we
obtain the probability for the charm or the anticharm quark to
hadronize by coalescence in the actual space-time volume
$\Delta t \Delta x \Delta y \Delta z$.  Whether the
charm or the anticharm quark is actually hadronized by coalescence is decided by
Monte Carlo.  Once the charm or the anticharm quark is decided to be
hadronized by coalescence, then we find its partner again by Monte
Carlo on the  basis on the probability of each combination in the
selected local ensemble. In case the charm or anticharm quark is
decided not to hadronize by coalescence, it is hadronized by the
fragmentation method as in p+p collisions (see Sec.~\ref{initial}). Since the
hadronization by coalescence is absent in p+p collisions, it can be
interpreted as a nuclear matter effect on the hadronization of charm and anticharm quarks.

\begin{figure}[h]
\centerline{
\includegraphics[width=9.5 cm]{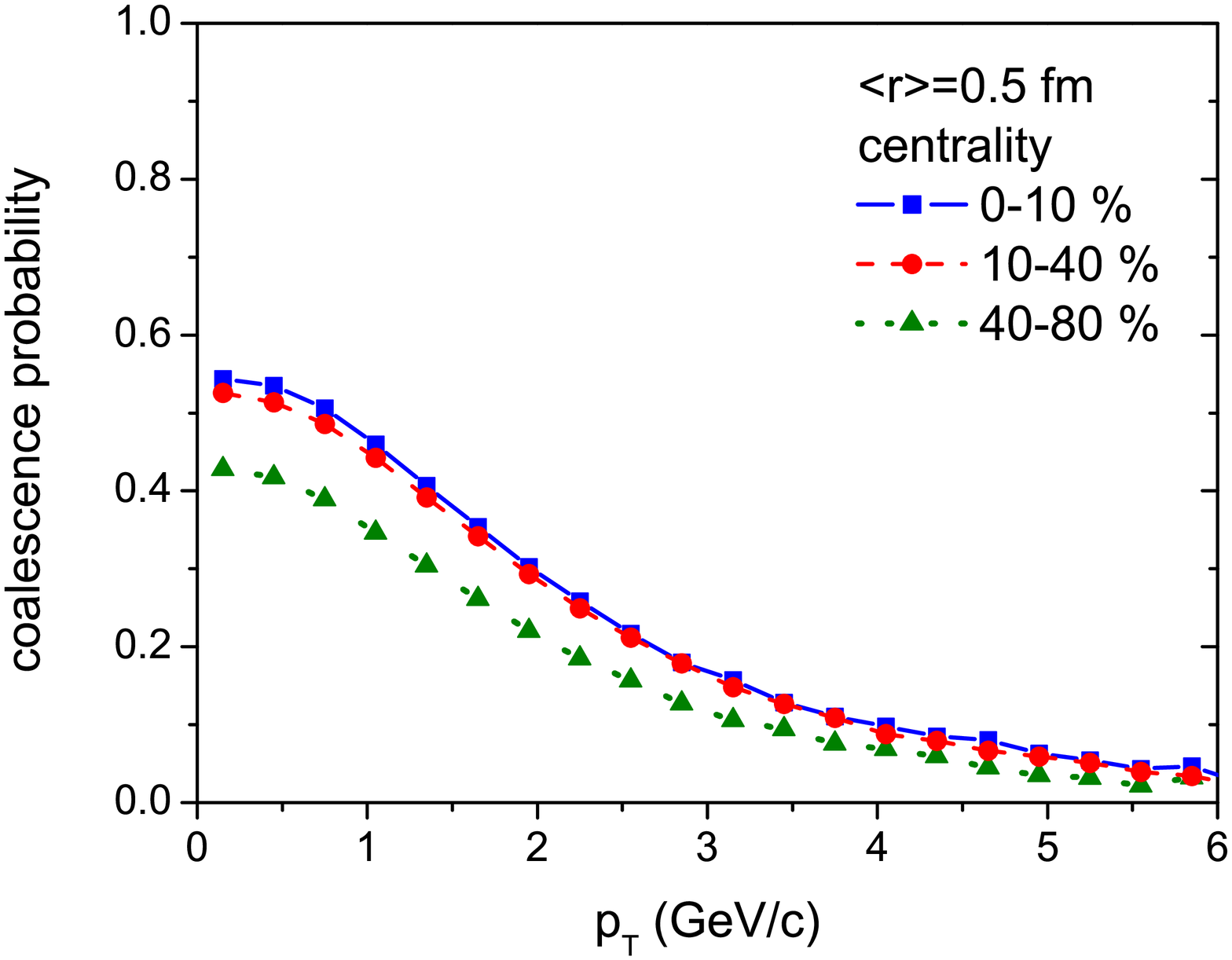}}
\centerline{
\includegraphics[width=9.5 cm]{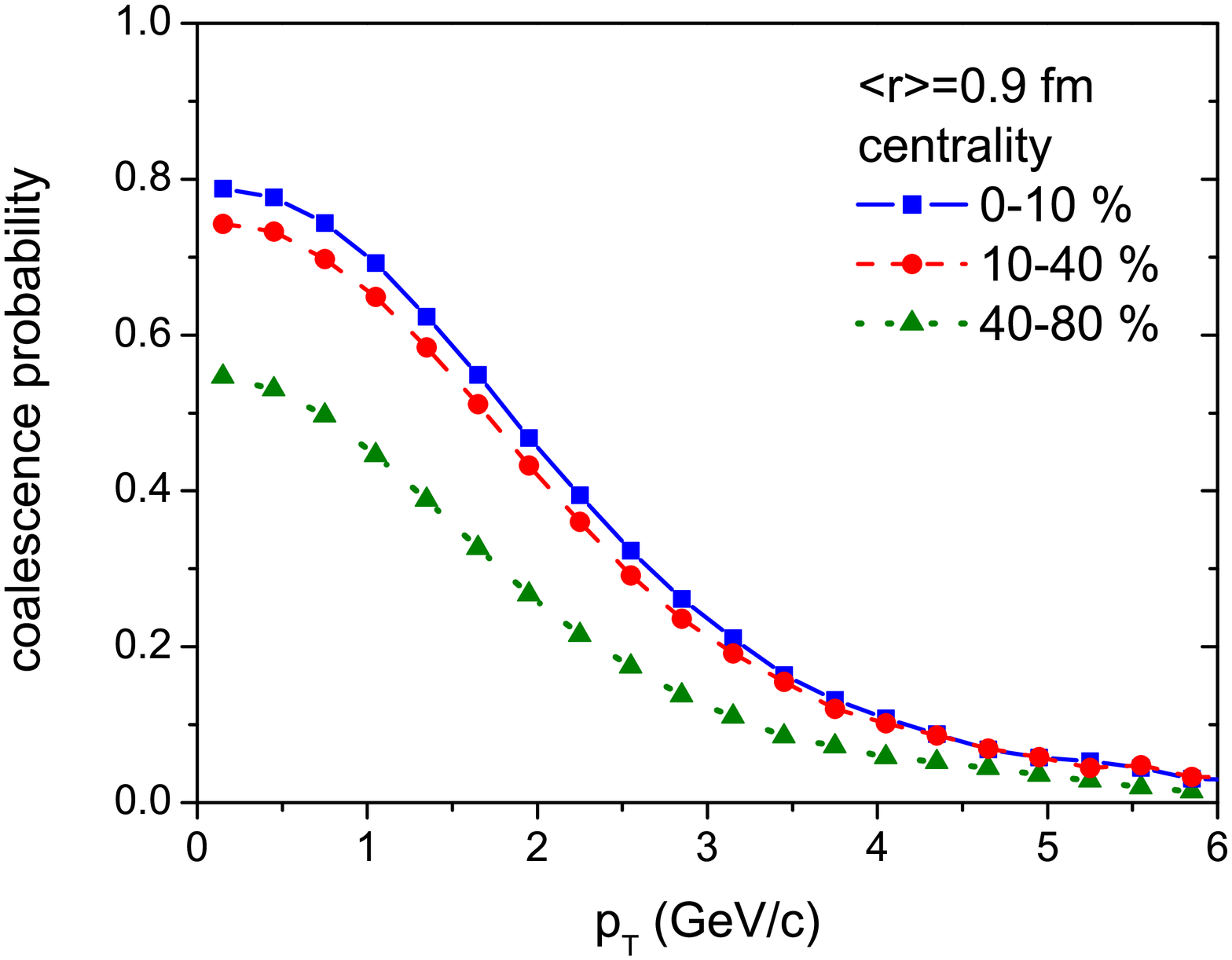}}
\caption{(Color online) The probabilities for charm or anticharm
quarks to hadronize to $D$ or $\bar{D}$ meson through
coalescence as a function of transverse momentum in 0-10,
10-40, 40-80 \% central Au+Au collisions at $\sqrt{s_{\rm NN}}=$200
GeV for the D meson radius of 0.5 fm (upper) and of 0.9 fm
(lower).} \label{probability}
\end{figure}

Fig.~\ref{probability} shows the probabilities for charm and
anticharm quarks to hadronize to $D$ and $\bar{D}$ mesons
through coalescence as a function of transverse momentum $p_T$ in
0-10, 10-40, 40-80 \% central Au+Au collisions at
$\sqrt{s_{\rm NN}}=$200 GeV for the D meson radius of 0.5 fm (upper) and
0.9 fm (lower). We can see that
the probability decreases as the transverse momentum of the charm or
anticharm quark increases. The reason is that the charm or anticharm quark
with larger transverse momentum has less chance to find a
neighboring coalescence partner in phase space.

Fig.~\ref{probability} also shows that coalescence probability
depends on centrality and the radius of the $D$ meson. Since there
are more light partons as a coalescence partner for the charm or
anticharm quark in central collisions, the coalescence probability
increases with decreasing centrality. In extremely peripheral
collisions, however, the coalescence probability becomes tiny -- as
in p+p collisions -- and the nuclear modification factor of $D$
mesons naturally approaches to 1. Since the hadronization of charm
quarks through fragmentation is not properly suited  at low
transverse momentum, several studies have forced the coalescence
probability to be 1 at
$p_T=$0~\cite{He:2011qa,Gossiaux:2010yx,Cao:2013ita,Oh:2009zj}.
Unless only a single centrality is concerned, it is technically
difficult to do so for all centralities by using the same
coalescence parameters. Moreover, if the coalescence probability is
1 in peripheral collisions, it will induce a large nuclear
modification factor, although small nuclear matter effect is
expected in such collisions. Therefore, we allow the fragmentation
of charm quarks at low transverse momentum as in p+p collisions in
competition with coalescence (for different radii of the
$D$-mesons). In Fig.~\ref{probability} we see that the coalescence
probability is larger for the $D$ meson radius of 0.9 fm than of 0.5
fm. We note that while the smaller radius is more physical for
$D$-mesons, a somewhat larger radius has been used in several
studies~\cite{Cao:2013ita,Oh:2009zj} to get a larger coalescence
probability.

The energy-momentum
difference between the charm or anticharm quark and the fragmented
$D$ or $\bar{D}$ meson in fragmentation and the energy
difference in coalescence are distributed equally to the
surrounding partons (in the same cell) which are not hadronized yet
to conserve the total energy-momentum of the system.

\section{$D$ meson scattering in the hadron gas}\label{HG2}

The $D$ and $D^*$ mesons produced through coalescence or
fragmentation interact with the surrounding hadrons in PHSD. The
presence of several resonances close to threshold energies with
dominant decay modes involving open-charm mesons and light hadrons
suggests that the scattering cross
sections of a $D/D^*$ off a meson or baryon, highly abundant in the
post-hadronization medium, could manifest a non-trivial energy,
isospin and flavor dependence. An example of these states is the
broad scalar resonance $D_0(2400)$, which decays into the
pseudoscalar ground state $D$ by emitting a pion in the $s$-wave
(similarly to the heavy-quark spin partner
$D_1(2420)$, decaying into $D^*\pi$\footnote{According
to quark model predictions, the excitation spectrum of the $D/D^*$
system consists of a triplet and a singlet of positive parity, with
$J^P=0^+,1^+,2^+$ and $1^+$, respectively (corresponding to
$^{2S+1}L_J=^3P_J$ and $^1P_1$). The $1^+$ states mix in such a way
that the meson with the lower mass, $D_1(2420)$, becomes narrow and
decouples from the $D^*\pi$ low-energy dynamics. The higher one,
$D_1(2420)$, on the contrary, is very broad and decays
predominantly in $D^*\pi$. The tensor state, $D_2^*(2460)$, is also
narrow and its influence can be, in principle, disregarded.}).
Moreover, the similarity  between the $\Lambda(1405)$ and the
$\Lambda_c(2595)$ has driven the attention to the fact that the
latter could be playing the role of a sub-threshold resonance in the
$DN$ system, connected to the latter by coupled-channel dynamics.

All these features have been addressed within several recent approaches based on hadronic effective models which incorporate chiral symmetry breaking in the light sector. The additional freedom stemming from the coupling to heavy-flavored mesons is constrained by imposing heavy-quark spin symmetry (HQSS) \cite{Abreu:2011ic,Abreu:2012et,Tolos:2013kva,Torres-Rincon:2014ffa,GarciaRecio:2008dp,Romanets:2012hm,Garcia-Recio:2013gaa}. Whereas chiral symmetry fully determines the scattering amplitudes of Goldstone bosons with other hadrons at leading order in a model independent way, by means of HQSS the dynamics of the pseudoscalar and the vector mesons containing heavy quarks can be connected, since all kinds of spin interactions are suppressed in the limit of infinite quark masses.

Following \cite{Abreu:2011ic} (and references therein), the Lagrangian density describing the interaction between the spin-zero and spin-one $D$ mesons with the light pseudoscalar (Goldstone) bosons from the octet ($\pi$, $K$, $\bar K$, $\eta$) reads $\cal L = \cal L_{\rm LO} + \cal L_{\rm NLO}$, where the subscripts LO and NLO refer to the leading and next-to-leading orders in the chiral perturbative expansion, whereas one keeps at leading order in the heavy-quark expansion. The LO Lagrangian reads
\begin{eqnarray}
\mathcal{L}_{LO} & = & \langle \nabla^\mu D \nabla_\mu D^\dag \rangle - m_D^2 \langle DD^{\dag} \rangle
- \langle \nabla^\mu D^{*\nu} \nabla_\mu D^{*\dag}_{\nu} \rangle \nonumber \\
&+& m_D^2 \langle D^{*\mu} D_\mu^{* \dag} \rangle
 + ig \langle D^{* \mu} u_\mu D^\dag - D u^\mu D_\mu^{*\dag} \rangle \nonumber \\
&+& \frac{g}{2m_D} \langle D^*_\mu u_\alpha \nabla_\beta D_\nu^{*\dag} -
\nabla_\beta D^*_\mu u_\alpha D_\nu^{*\dag}\rangle  \epsilon^{\mu \nu \alpha\beta}\ ,
\end{eqnarray}
where $D=(D^0,D^+,D_s^+)$  and $D_{\mu}^*=(D^{*0},D^{*+},D_s^{*+})_{\mu}$ are the SU(3) antitriplets of spin-zero and spin-one $D$ mesons, respectively, with mass $m_D$ in the chiral limit, and the brackets denote the trace in flavor space.
Two interaction terms are present in the LO Lagrangian, containing the coupling of the $D$ and $D^*$ meson fields to the light-meson axial vector current, $u_\mu = i \left( u^\dag \partial_\mu u - u \partial_\mu u^\dag \right)$, which describe, for instance, the decay of the $D^*$ meson into a $D\pi$ pair. We note that the two terms are in principle independent but HQSS demands that both are related by a unique coupling constant $g$.
The covariant derivative contains the coupling to the light-meson vector current and reads $\nabla_\mu = \partial_\mu - \frac{1}{2} \left( u^\dag \partial_\mu u + u \partial_\mu u^\dag \right)$, where the Goldstone bosons are introduced within the non-linear realization of chiral symmetry in exponential parameterization, $U=u^2=\exp \left(\frac{\sqrt{2} i\Phi}{f}  \right)$, with
\begin{equation}
\Phi= \left(
\begin{array}{ccc}
\frac{1}{\sqrt{2}} \pi^0 + \frac{1}{\sqrt{6}} \eta  & \pi^+ & K^+ \\
\pi^- & - \frac{1}{\sqrt{2}} \pi^0 + \frac{1}{\sqrt{6}} \eta & K^0 \\
K^- & {\bar K}^0 & - \frac{2}{\sqrt{6}} \eta
\end{array}
\right)
\end{equation}
and $f$ the meson decay constant in the chiral limit ($f=93$~MeV).

The NLO Lagrangian, which we omit here for simplicity, introduces twelve low-energy constants (LECs), $h_i$ and $\tilde{h}_i$ $(i=0,\dots,5)$, which have to be fixed by symmetry arguments and known phenomenology. Imposing HQSS at LO reduces the number of free parameters to six, since $h_i=\tilde{h}_i$, whereas by large-$N_c$ considerations it can be argued that only the odd LECs contribute.

\begin{figure*}[t]
\centerline{
\includegraphics[width=9.5 cm]{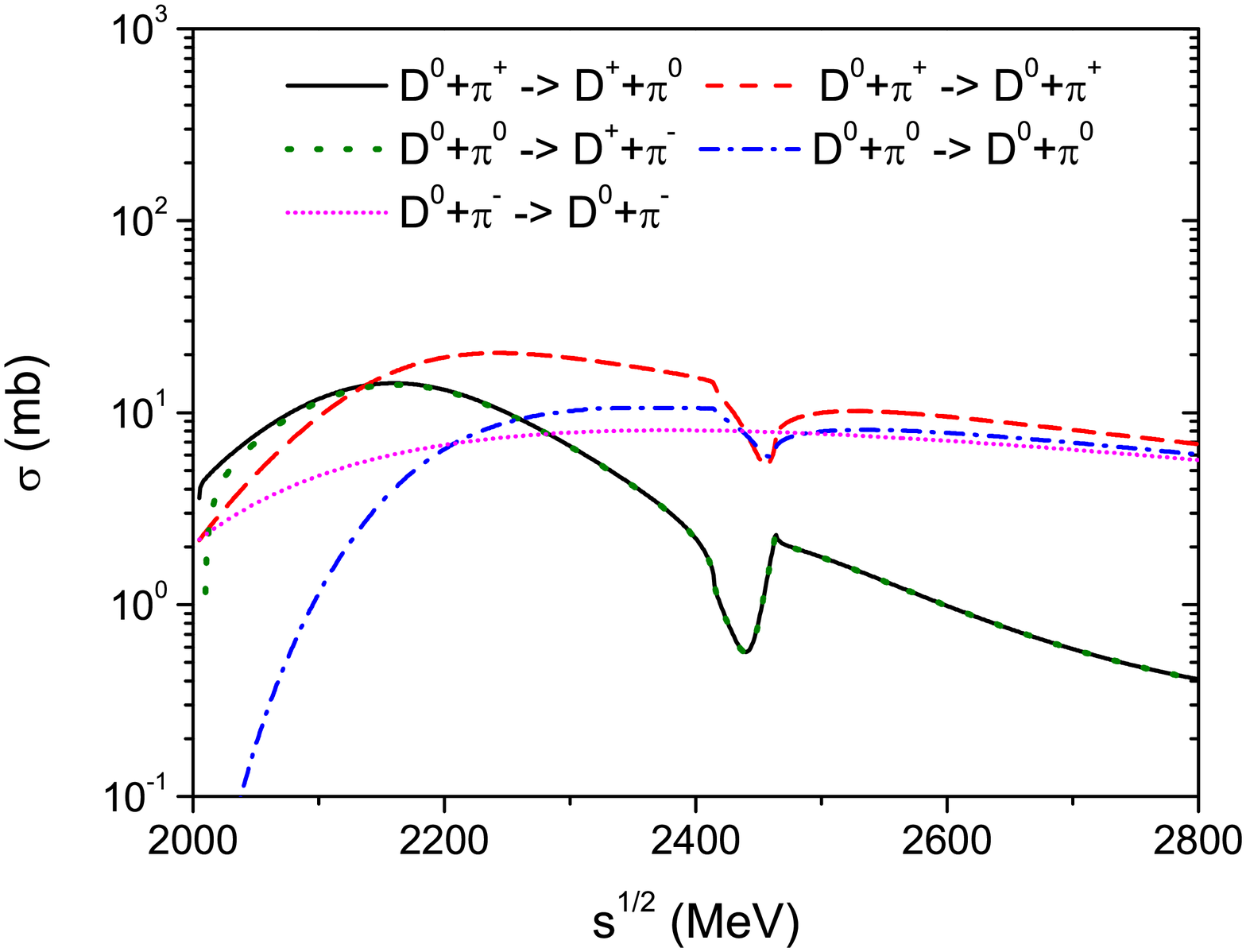}
\includegraphics[width=9.5 cm]{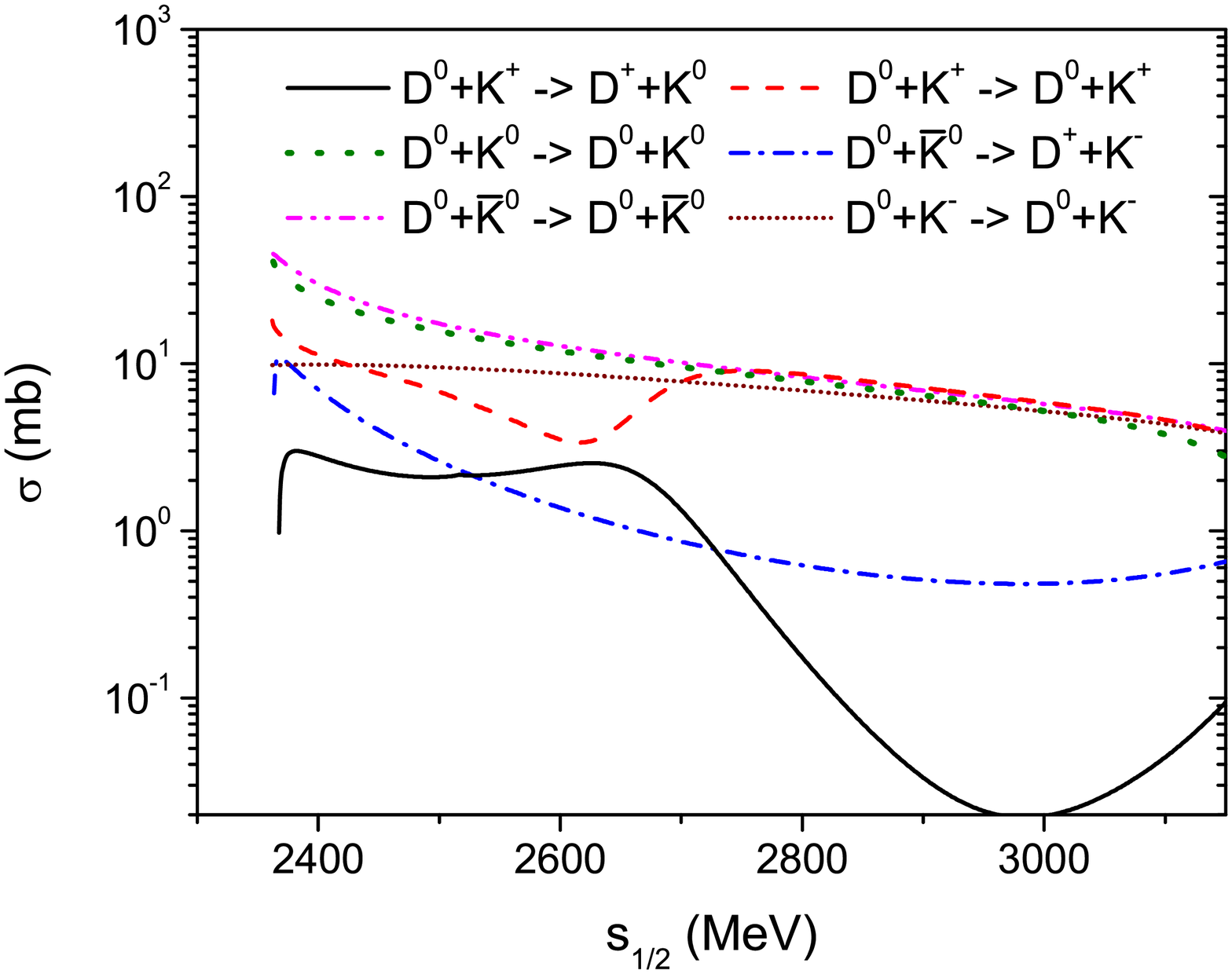}}
\centerline{
\includegraphics[width=9.5 cm]{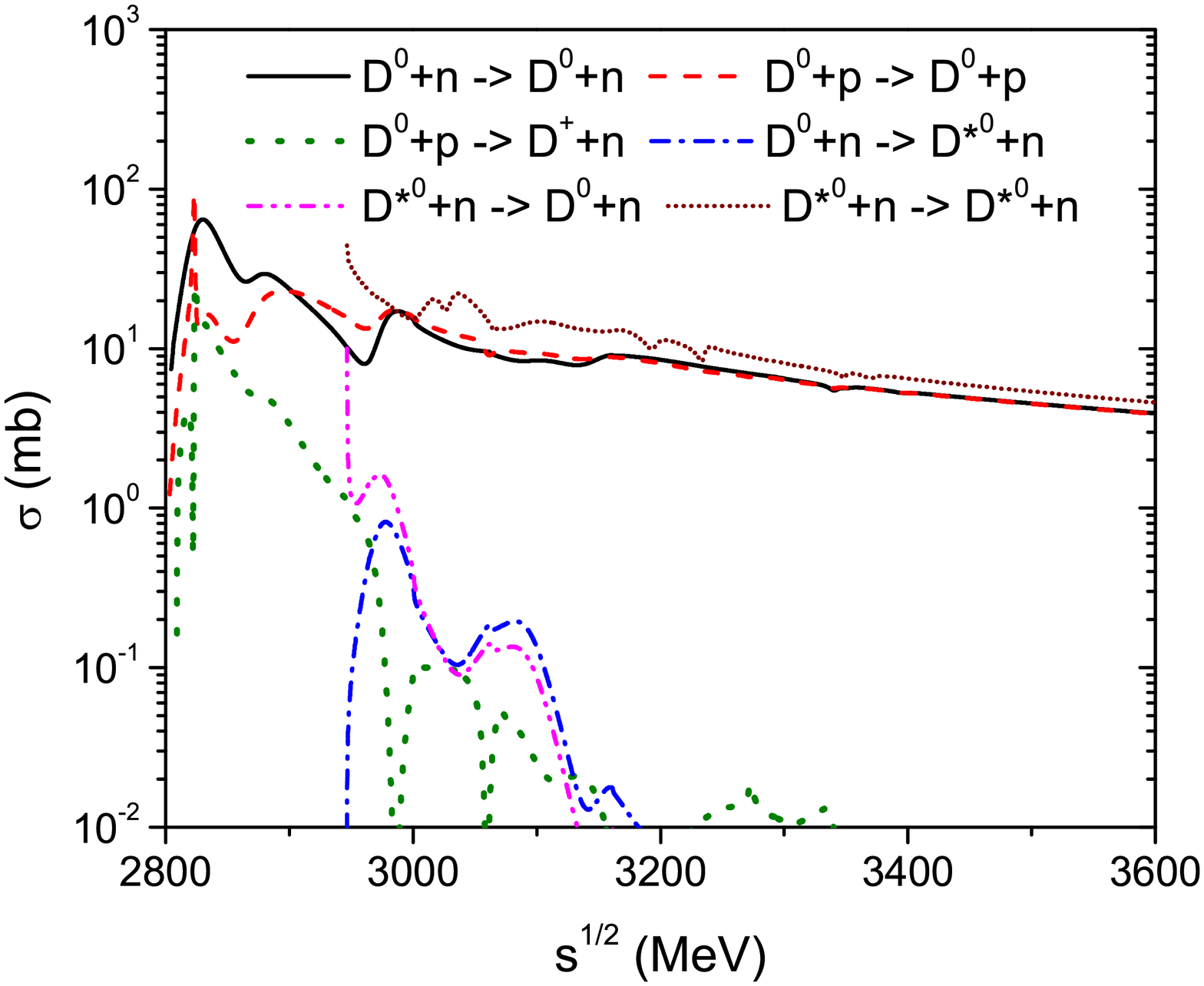}
\includegraphics[width=9.5 cm]{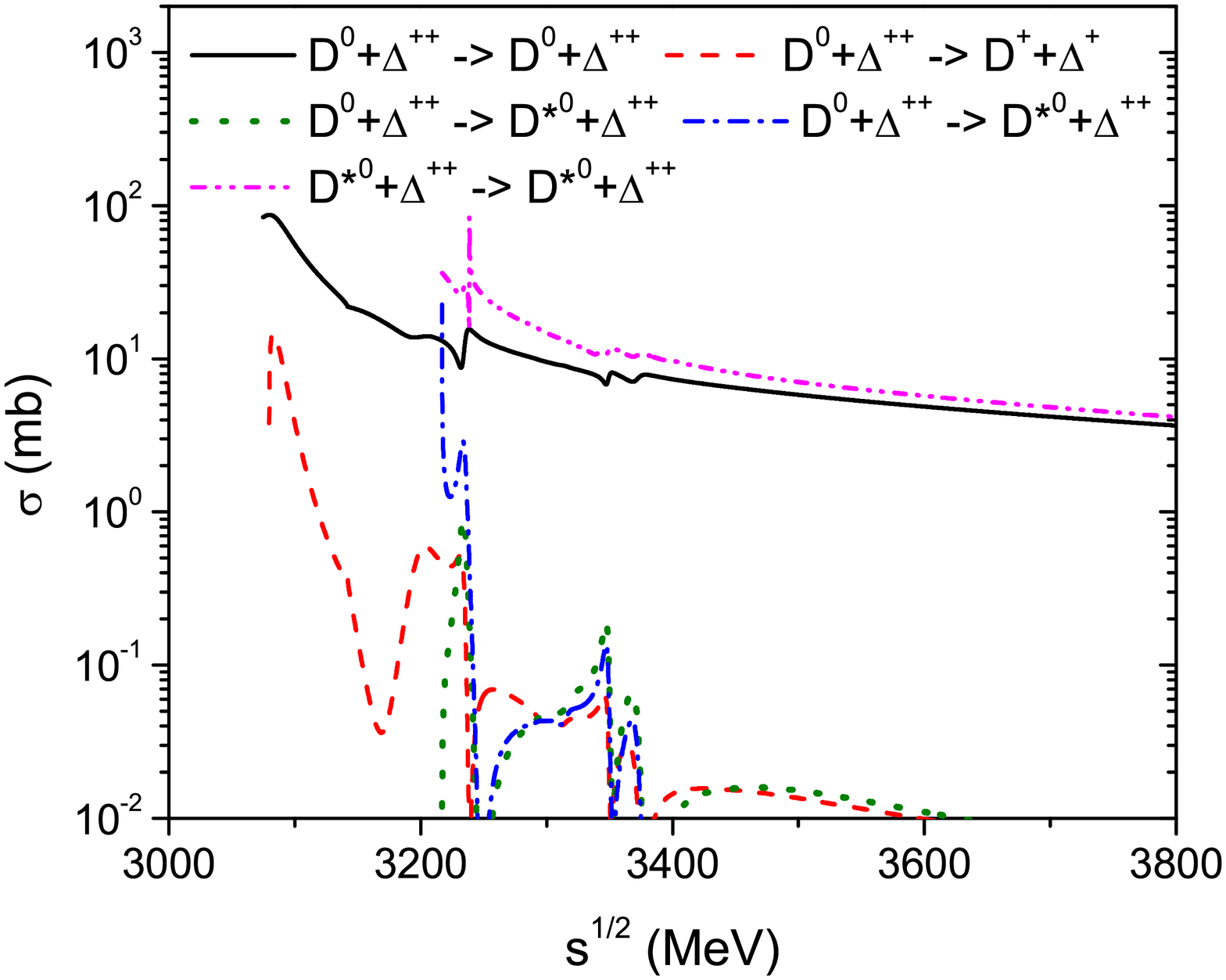}}
\caption{(Color online) Several examples for the scattering cross sections of $D$ and $D^*$ mesons with light mesons and baryons.}
\label{HG3}
\end{figure*}

The interaction of $D$ mesons with baryons has been recently studied in different approaches which incorporate heavy flavor in the meson baryon interaction, either by implementing $t$-channel vector-meson exchange mechanisms between pseudoscalar mesons and baryons, by extending the J\"ulich meson-exchange model, or relying on the hidden gauge formalism (see \cite{Tolos:2013gta} and references therein). Among these we recourse to the model described in Refs.~\cite{GarciaRecio:2008dp,Romanets:2012hm,Garcia-Recio:2013gaa,Gamermann:2011mq}, which generalizes the Weinberg-Tomozawa form of the meson-baryon interaction, fixed by chiral symmetry at leading order, beyond flavor SU(3) structure. In addition, in this model HQSS is fulfilled exactly whenever charm quarks participate in the interaction.
In all charm sectors, and in particular in $C=1$ ($C=$~charm number), the theory accounts for point-like $s$-wave interactions between charmed mesons and light mesons in the pseudoscalar and vector octets, and the lowest lying baryons from the $J^P=1/2^+$ (nucleon) octet and the $3/2^+$ (Delta) decuplet. It reduces to the WT interaction prescribed by chiral symmetry as long as the pseudoscalar mesons (Goldstone bosons) are involved. The latter is accomplished by enhancing the standard SU(3) flavor symmetry to a SU(6) spin-flavor symmetry [SU(8) when charm is also considered], which has allowed to identify unambiguously many baryonic states as dynamically generated resonances of the meson-baryon interaction in the light sector \cite{GarciaRecio:2008dp,Romanets:2012hm,Garcia-Recio:2013gaa,Gamermann:2011mq}.
HQSS imposes that arbitrary rotations of the heavy-quark spin should leave dynamics unchanged, which prevents charm flavor to be exchanged between the interacting meson-baryon pair, naturally leading to the
suppression of charm-exchange processes\footnote{This feature arises in other models based on meson exchange mechanisms due to the heavy mass of the charmed field and the large heavy-meson decay constants; note, however, that HQSS is not exactly fulfilled in most of these models \cite{Garcia-Recio:2013gaa}.}.
The tree-level amplitudes of the theory are given by
\begin{eqnarray}
V_{ij}^{IJSC}(\sqrt{s}) &=& D_{ij}^{IJSC} \, \frac{2\sqrt{s}-M_i-M_j}{4 f_i f_j} \nonumber \\
&& \times \sqrt{\frac{E_i+M_i}{2 M_i}} \sqrt{\frac{E_j+M_j}{2 M_j}} \ ,
\label{eq:V-spin-flavor}
\end{eqnarray}
where $M_i$ and $E_i$ denote the mass and CM energy of the baryon in channel $i$, respectively, $f_i$ stands for the corresponding meson decay constant, and $D^{IJSC}$ is a matrix of coefficients in the coupled-channel space for given isospin, spin, strangeness and charm numbers. Breaking of the SU(8) symmetry is taken into account by using physical hadron masses and meson decay constants in Eq.~(\ref{eq:V-spin-flavor}).
The model is not complete in the analytic sense since no $t$- or $u$- channel mechanisms are explored; it is, however, a minimal extension of the SU(3) WT interaction with no additional free parameters which simultaneously implements the relevant symmetries of QCD in the light- and heavy-quark sectors.

The tree-level scattering amplitudes obtained from the previous approaches to the scattering of $D/D^*$ mesons with light mesons and baryons are unitarized along the right-hand cut by solving the set of coupled-channel (on-shell) Bethe-Salpeter equations, $T=T+VGT$. Here $T$~($V$) stands for the unitarized~(tree-level) amplitude with matrix element $T$($V$)$^{IJSC}_{ij}$, where $i(j)$ labels the incident~(outgoing) heavy-light meson or heavy-meson baryon state. $G$ represents the two-particle propagator or loop function, which is regularized by subtractions or in dimensional regularization.
The few parameters unconstrained in the heavy-light meson Lagrangian are  fixed by the mass difference between the $D$ and $D_s$ mesons, and by the position and width of the $D_0^*(2400)$ resonance, which is dynamically generated by the unitarized interaction in the $I=1/2$ $D\pi$ channel. We recall that the model for the meson-baryon interaction is parameter free, modulo regularization.
The sub-threshold $\Lambda_c(2595)$ resonance is dynamically generated by the model, with a strong coupling to the $DN$ and $D^*N$ channels in $I=0$ and a finite width from the $\Sigma_c\pi$ channel, which is open for decay. Several additional states in the $C=1$ and $S=0$ sector are generated by the interaction, some of which have been identified with experimental observations whereas others are genuine predictions that could be detected in the forthcoming CBM and PANDA experiments at GSI/FAIR \cite{Romanets:2012hm}.

A selection of the $D$ and $D^*$ cross sections off light mesons, nucleons and $\Delta$ baryons is shown in Fig.~\ref{HG3} for several physical (charge) states. The cross sections exhibit a remarkable isospin dependence, as expected, leading to rather different shapes (energy dependence). For example, the $D_0^*(2400)$ resonance is clearly visible in the $D^0\pi^+$ elastic reaction, which has a predominant $I=1/2$ component. Conversely, the $D^0\pi^-$ elastic reaction is pure $I=3/2$ and the cross section is not resonant in this case, whereas other processes exhibit deviations from the resonance profile due to the interference between the $I=1/2,3/2$ amplitudes. We note the dip caused by the opening of the $D\eta$ channel, which mixes with $D\pi$ only at NLO in the chiral Lagrangian.
The $D\bar K$ cross sections present a characteristic monotonic fall due to the presence of the narrow $D_s^*(2317)$, which lies right below threshold energy (compare, e.g, with the cross section of the pure isovector $D^0 K^-$, which is essentially flat). The isoscalar part of the $DK$ amplitude contributes with a broad resonant state around $2700$~MeV. Significantly, the vector partner $D_{s1}^*(2700)$ has been observed with a decay width of about $120$~MeV. Similar cross sections are obtained for the $D^*$ interaction as the underlying dynamics is identical at LO in the heavy-quark expansion, and the only differences are introduced by the slightly higher mass of the $D^*$ (thus thresholds and resonances appear shifted by this amount).
In addition to the subthreshold $\Lambda_c$ and $\Sigma_c$ states generated from the $D$ and $D^*$ interactions with baryons, the $D^{(*)}N$ and $D^{(*)}\Delta$ cross sections are populated by several resonances at higher energies, with a remarkable isospin and coupled-channel dependence. In both cases the reactions with a strong isovector component seem to dominate over other processes. We also account explicitly for the cross sections of reactions with antinucleons, not shown in Fig.~\ref{HG3}, which are obtained from the $\bar D N$ scattering amplitudes by charge conjugation.
Finally, the cross sections for the scattering of $D$ and $D^*$ with the vector mesons from the octet (e.g. $D\rho$), out of the scope of the present model, have been set to an estimated value of $10$~mb and independent of the collision energy.

\begin{figure}[h!]
\centerline{
\includegraphics[width=9.5 cm]{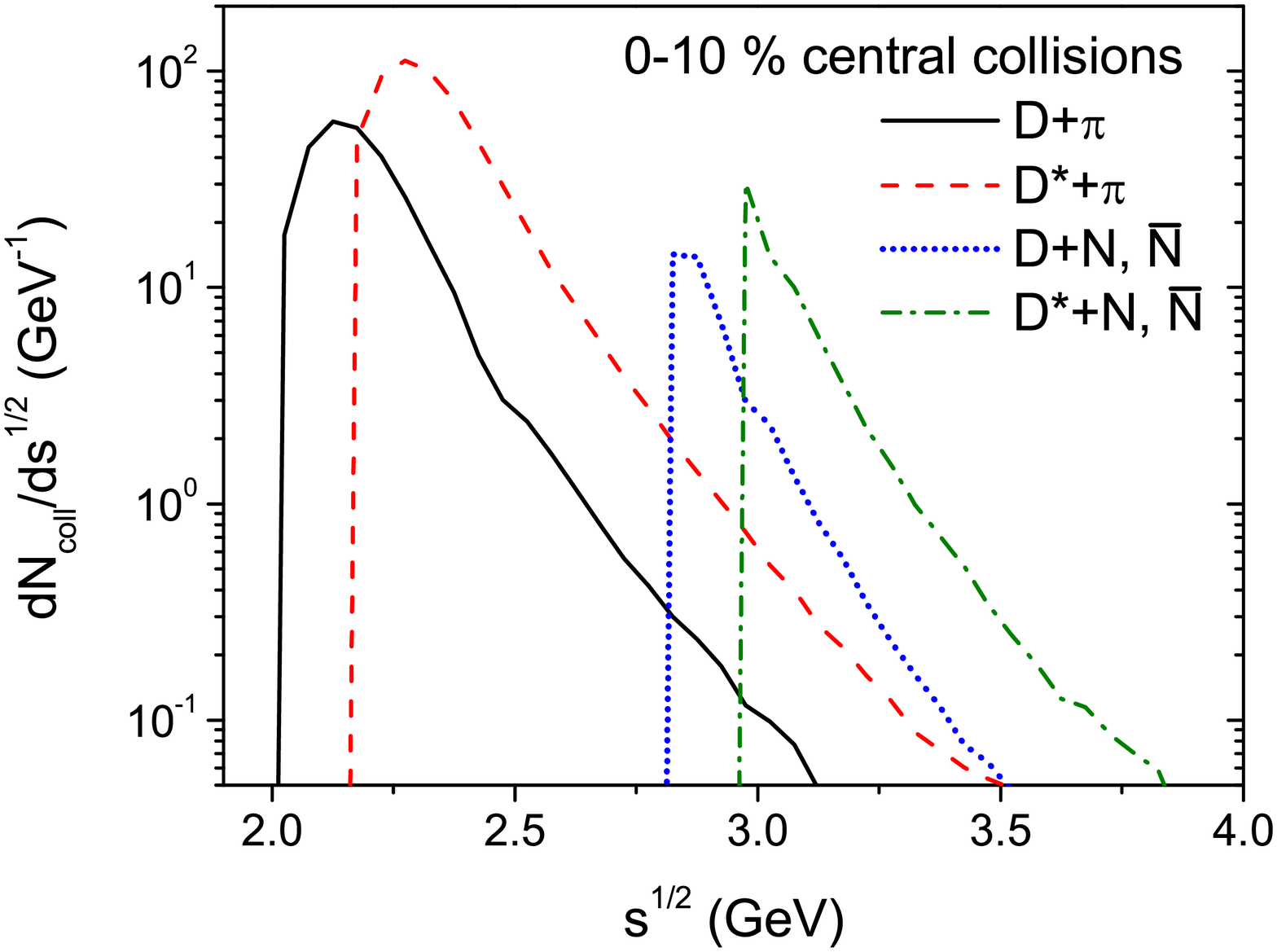}}
\caption{(Color online) The distributions of $D$ and $D^*$ meson scattering
with pions and nucleons as a function of the invariant scattering energy $\sqrt{s}$
in 0-10 \% Au+Au central collisions. See the legend for the individual channels.}
\label{scatterH}
\end{figure}

Fig.~\ref{scatterH} shows the distribution of $D$ and $D^*$
meson scatterings with pions and nucleons as a function of the invariant
scattering energy $\sqrt{s}$  in 0-10 \% Au+Au central collisions. The total number of
$D$ and $D^*$ meson scatterings with pions is about 49 and that with
nucleons or antinucleons is $\sim$ 6. The numbers of $D$ and $D^*$ meson
scatterings with all mesons and all baryons are, respectively, $\sim$ 56
and  $\sim$ 10. Considering that the number of $D$ and $D^*$ mesons in
0-10 \% central collisions is about 30, excluding $D_s$, $D_s^*$,
$\Lambda_c$, and $\bar{\Lambda}_c$, each $D$ or $D^*$ meson
experiences on average two scatterings with a hadron until it
freezes out. Compared to Fig.~\ref{scatterP}, the number of
scatterings decreases rapidly with increasing scattering energy.
This is attributed to the decrease of hadronic cross sections beyond the resonance region, as shown in Fig.~\ref{HG3}.
Accordingly,  hadronic interactions become
 ineffective for the energy loss of
$D$ and $D^*$ mesons at high transverse momentum.

\section{results}\label{results}
\begin{figure}[tbh]
\centerline{
\includegraphics[width=9.5 cm]{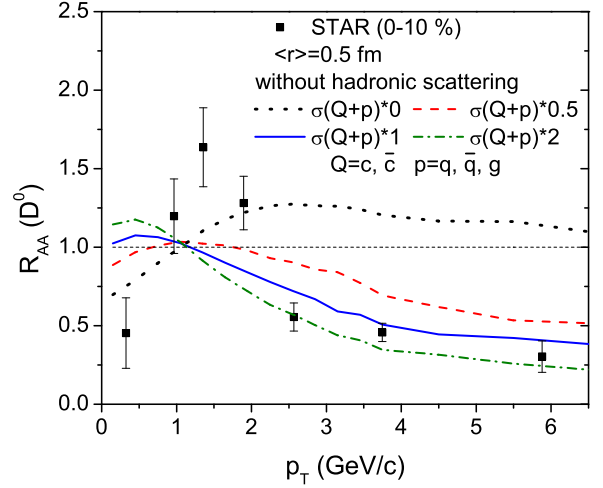}}
\centerline{
\includegraphics[width=9.5 cm]{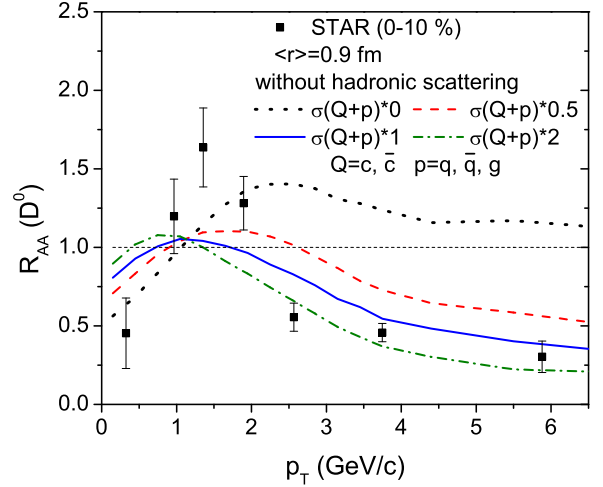}}
\caption{(Color online) The $\rm R_{AA}$ of $D^0$ mesons without
hadronic scattering in 0-10 \% central Au+Au collisions at
$\sqrt{s_{\rm NN}}=$200 GeV for a radius of the $D$ meson of 0.5 fm (upper) and
of 0.9 fm (lower), which are compared with
the experimental data from the STAR
Collaboration~\cite{Adamczyk:2014uip}. The cross sections for
charm-parton scattering are multiplied by factors of 0, 0.5, 1, and
2 for the dotted, dashed, solid, and dot-dashed lines,
respectively.} \label{partons}
\end{figure}

The nuclear modification of $D$ mesons is expressed in term of the ratio $\rm R_{AA}$ which is defined as
\begin{eqnarray}
{\rm R_{AA}}(p_T)\equiv\frac{dN_D^{\rm Au+Au}/dp_T}{N_{\rm binary}^{\rm Au+Au}\times dN_D^{\rm p+p}/dp_T},
\label{raa}
\end{eqnarray}
where $N_D^{\rm Au+Au}$ and $N_D^{\rm Au+Au}$ are, respectively, the number of
$D$ mesons produced in Au+Au collisions and that in p+p collisions, and
$N_{\rm binary}^{\rm Au+Au}$ is the number of binary nucleon-nucleon
collisions in Au+Au collision for the centrality class considered. If
the matter produced in relativistic heavy-ion collisions does not
modify the $D$ meson production and propagation, the numerator of
Eq.~(\ref{raa}) should be equal to the denominator. Therefore, an $\rm
R_{AA}$ smaller or larger than one implies that the nuclear matter
suppresses or enhances $D$ mesons, respectively.

\begin{figure}[tbh]
\centerline{
\includegraphics[width=9.5 cm]{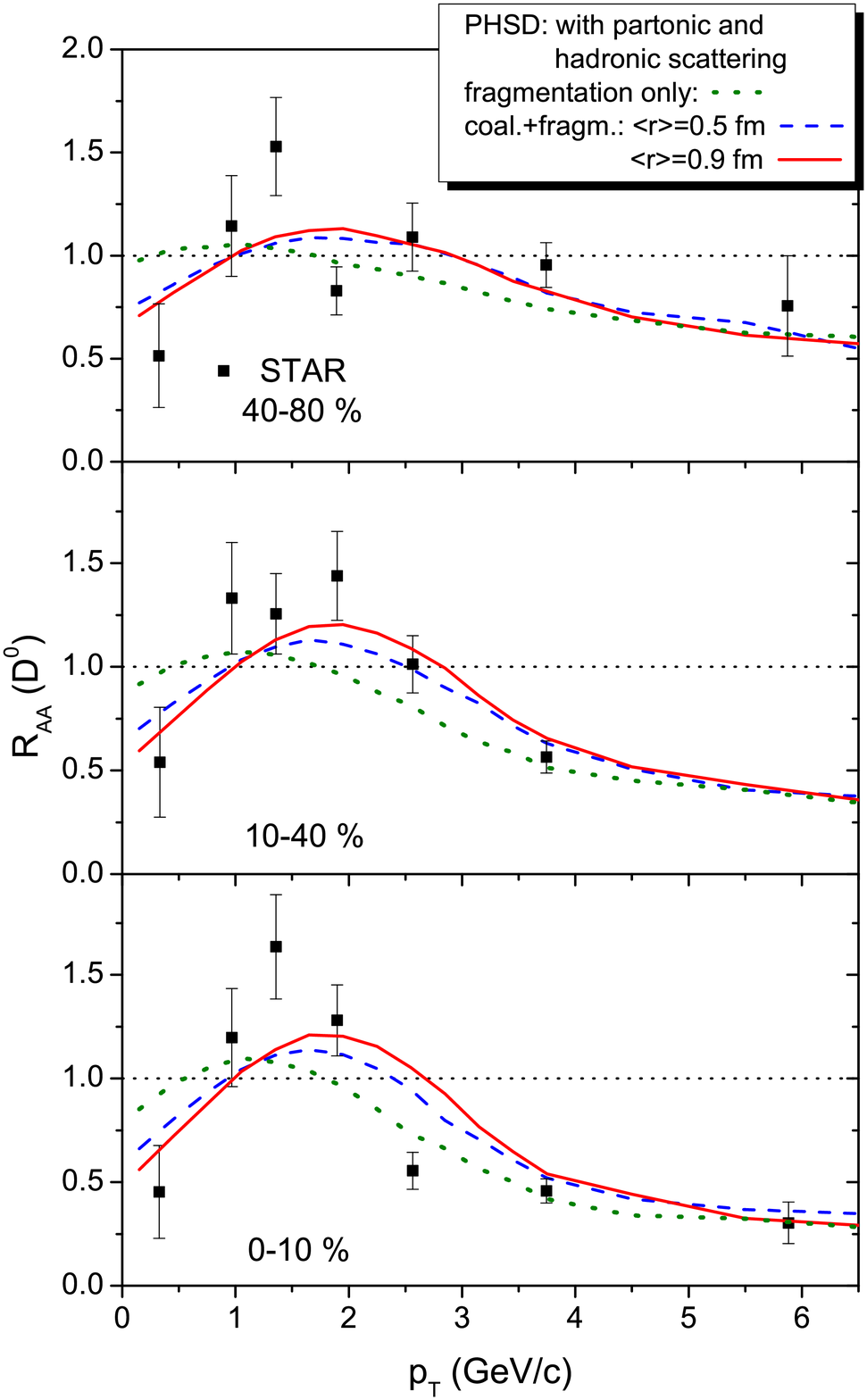}}
\caption{(Color online) The $\rm R_{AA}$ of $D^0$ mesons in Au+Au collisions at $\sqrt{s_{\rm
NN}}=$200 GeV including partonic and hadronic scatterings for the $D$ meson radius of 0.5 fm (dashed lines), of 0.9 fm (solid lines), and without coalescence, i.e. fragmentation only (dotted lines), which are compared with the experimental data from the STAR
Collaboration~\cite{Adamczyk:2014uip}.}
\label{coalescence1}
\end{figure}

In order to see the effect of partonic scattering, all hadronic
scatterings of $D$ mesons are switched off in Fig.~\ref{partons}.
The dotted, dashed, solid, and dot-dashed lines in the figure show
the $\rm R_{AA}$ of $D^0$ mesons in 0-10 \% central Au+Au collisions
with the cross sections for the partonic scattering of charm being
artificially multiplied by factors of 0, 0.5, 1, and 2, respectively.
We stress that these multiplication factors are introduced in order to explore the impact of the
partonic interaction strength on the shape of $R_{AA}(p_T)$. Note, however, that the default cross
section (multiplication factor 1) is determined consistently on the basis of the DQPM couplings and propagators and thus has no free parameters. The fact that the $R_{AA}(p_T)$ is best described by the consistent cross sections (blue solid lines) points towards an experimental support of our approach.
The black dotted line, where both partonic and hadronic scatterings are
absent, shows that $\rm R_{AA}$ approches 1 at high $p_T$ as it
ought to be. However, the coalescence of charm quarks creates a peak
in $\rm R_{AA}$ around $p_T=$ 2.5 GeV/c. This peak does not show up in
the PHSD calculations when discarding the dissolution of strings to
partons. In this case the $\rm R_{AA}$ is unity. The peak emerges because the charm hadron gains
transverse momentum in the coalescence which is absent in p+p
collisions. The red dashed, blue solid, and green dot-dashed lines
show that with increasing cross section for charm and parton
scattering the charm quark looses more energy at high $p_T$. It is
also seen that the energy loss of a charm or anticharm quark at high
$p_T$ can be dominantly attributed to the interaction with partons in
the QGP. On the other hand, the $\rm R_{AA}$ at low $p_T$ increases
with increasing scattering cross section. The reason is that a
larger scattering cross section produces charm quarks closer to
their thermal equilibrium distribution as shown in
Fig.~\ref{pt-QGP}, which is enhanced at low $p_T$ relative to the
initial $p_T$ distribution.

\begin{figure}[tbh]
\centerline{
\includegraphics[width=9.5 cm]{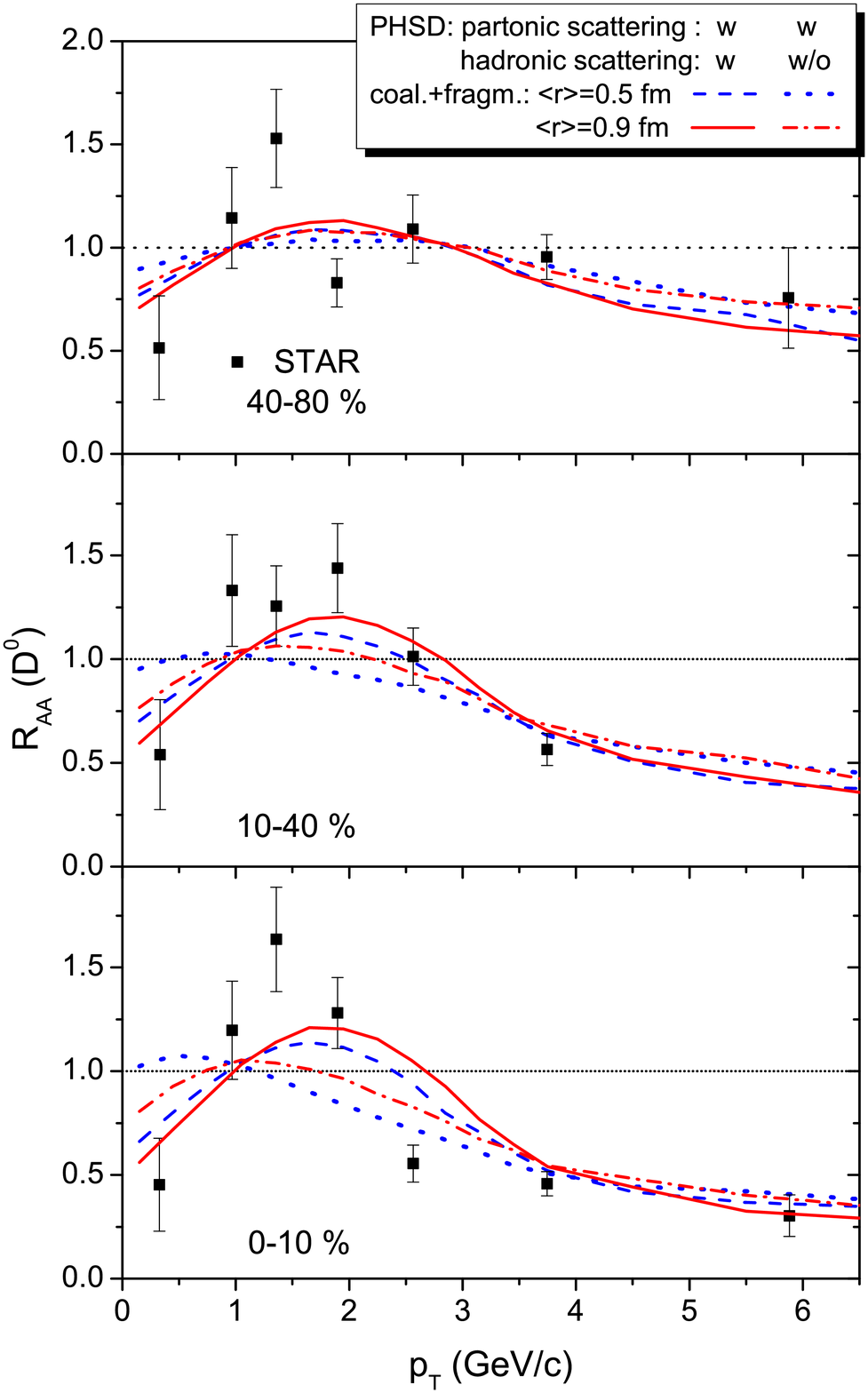}}
\caption{(Color online) The $\rm R_{AA}$  of
$D^0$ mesons including partonic scattering with (dashed and solid lines) and without hadronic scattering (dotted and dot-dashed lines)
in Au+Au collisions at $\sqrt{s_{\rm NN}}=$200 GeV for a $D$ meson
radius of 0.5 fm and of 0.9 fm. The experimental data are from the
STAR Collaboration~\cite{Adamczyk:2014uip,Tlusty:2012ix} .}
\label{hadrons}
\end{figure}

\begin{figure}[tbh]
\centerline{
\includegraphics[width=9.5 cm]{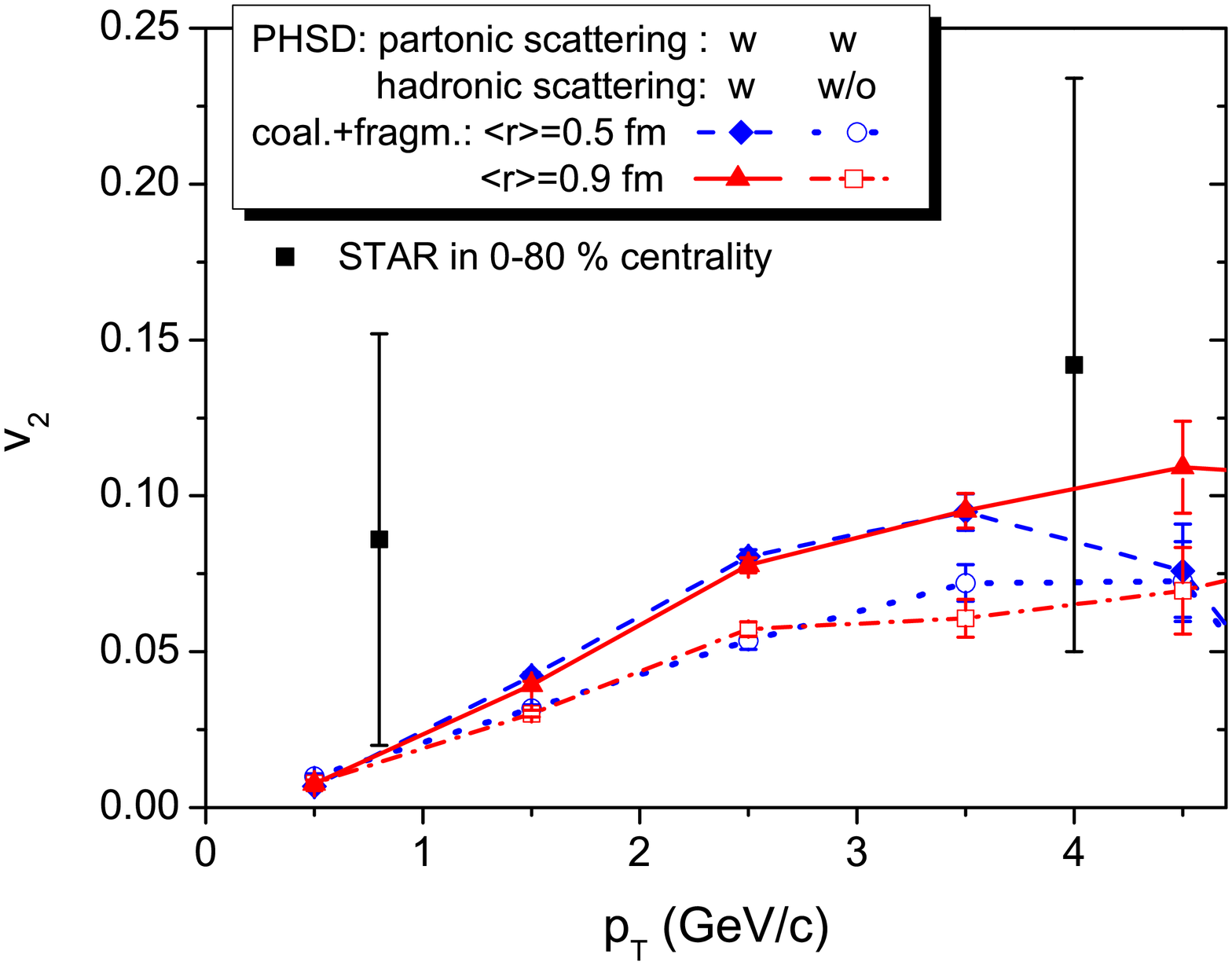}}
\caption{(Color online) The elliptic flow $v_2$ of $D^0$ mesons including partonic scattering with
(dashed and solid lines) and without
hadronic scattering (dotted and
dot-dashed lines) in Au+Au collisions at
$\sqrt{s_{\rm NN}}=$200 GeV for a $D$ meson radius of 0.5
fm and of 0.9 fm. The experimental data are from the STAR
Collaboration~\cite{Adamczyk:2014uip,Tlusty:2012ix}.}
\label{hadrons2}
\end{figure}

In Fig.~\ref{coalescence1} we show the effect of the coalescence
probability on the $\rm R_{AA}$ of $D^0$ mesons. The dashed and solid
lines are, respectively, the $\rm R_{AA}$ for a $D$ meson radius of 0.5 fm and
of 0.9 fm as shown in Fig.~\ref{probability}. For the
dotted lines, all charm and anticharm
quarks are hadronized by fragmentation. It is seen that
for the $D$ meson radius
of 0.9 fm more charm quarks are hadronized through coalescence
and as a result the peak of $\rm R_{AA}$ is
shifted to higher $p_T$. On the contrary, when all charm quarks are
hadronized through fragmentation, the peak is shifted to lower
$p_T$.

Finally, Fig.~\ref{hadrons} shows the $\rm R_{AA}$  and Fig.~\ref{hadrons2} the elliptic flow $v_2$ of
$D^0$ mesons  with and without hadronic scattering. We can see that the
hadronic scattering plays an important role both in $\rm R_{AA}$ and
the elliptc flow $v_2$. It shifts the peak of $\rm R_{AA}$ to higher
transverse momentum especially in central collisions and enhances the
elliptic flow of final $D$ mesons.

\section{summary}\label{summary}

We have studied charm production in relativistic heavy-ion
collisions by using the Parton-Hadron-String Dynamics (PHSD) approach
\cite{PHSDrhic} where the initial charm quark pairs are produced in binary
nucleon-nucleon collisions from the PYTHIA event generator \cite{Sjostrand:2006za} taking into
account the smearing of the collision energy due to the Fermi motion of
nucleons in the initial nuclei.  The produced charm and anticharm
quarks interact with the dressed quarks and gluons in the QGP which are
described by the Dynamical Quasi-Particle Model \cite{Cassing:2008nn} in PHSD. The interactions of
the charm quarks with the QGP partons have been evaluated with the DQPM
propagators and couplings consistently~\cite{Berrehrah:2013mua}.
Furthermore, when extracting the spatial diffusion coefficient $D_s$ from our
cross sections (Fig. \ref{fig:cDs}) as a function of the temperature we observe
a minimum of $D_s$ close to $T_c$ which is in line with lattice data above $T_c$
and hadronic many-body calculations below $T_c$.

We recall that the PHSD differs from conventional Boltzmann
approaches incorporating on-shell scattering with pQCD cross
sections in a couple of essential aspects:\\  i) it incorporates
dynamical quasi-particles due to the finite width of the spectral
functions; \\ ii) it involves scalar mean-fields for the light
partons that substantially drive the collective flow in the partonic
phase and includes $c$-quark scattering with the QGP partons that transfers
collective flow also to the charm quarks;\\  iii) it is based on a
realistic equation of state from lattice QCD and thus describes the
speed of sound $c_s(T)$ reliably (without incorporating a first
order phase transition);\\  iv) the hadronization of 'bulk' partons
is described by the fusion of off-shell partons to off-shell
hadronic states (resonances or strings) and does not violate the
second law of thermodynamics; \\ v) all conservation laws
(energy-momentum, flavor currents etc.) are fulfilled in the
hadronization (contrary to some coalescence models);\\  vi) the
effective partonic cross sections no longer are given by pQCD and
are evaluated within the DQPM in a consistent fashion and probed by
transport coefficients (correlators) in thermodynamic equilibrium
(shear- and bulk viscosity, electric conductivity, magnetic
susceptibility, spatial charm diffusion coefficient etc.
\cite{Hamza14,Vitaly2,Ca13}).

We have found that the
interaction with the dynamical partons of the QGP softens the $p_T$ spectrum of charm and
anticharm quarks but does not lead to a full thermalization for transverse
momenta $p_T >$ 2 GeV/c.  The charm
and anticharm quarks, furthermore,  are hadronized to $D$ mesons either through the
coalescence with a light quark or antiquark or through the
fragmentation by emitting soft 'perturbative' gluons.
Since the hadronization through
coalescence is absent in p+p collisions, it can be interpreted as a
nuclear matter effect on the $D$ meson production in relativistic
heavy-ion collisions.  In the coalescence mechanism the charm or
anticharm quark gains momentum by fusing with a light quark or
antiquark while it looses momentum in the fragmentation process (as in
p+p reactions).  This partly contributes to the large $\rm R_{AA}$ of
$D$ mesons between 1 and 2 GeV/c of transverse momentum.  Finally, the
formed $D$ mesons interact with hadrons by using the cross sections
calculated in an effective lagrangian approach with heavy quark
spin-symmetry \cite{Tolos:2013kva}, which is state-of-the art.  We have found that the
contribution from hadronic scattering both to the $\rm R_{AA}$ and to
the elliptic flow of $D$ mesons is appreciable, especially in central
collisions, and produces additional elliptic flow $v_2$.

Since the PHSD results reproduce the experimental data from the STAR
Collaboration without radiative energy loss in the $p_T$ range
considered, we conclude that collisional energy loss is dominant at
least up to $p_T=$ 6 GeV/c in relativistic heavy-ion collisions. In our
approach this is essentially due to the infrared enhanced coupling $\alpha_s(T)$ in
the DQPM leading to large scattering cross sections of charm quarks
with partons at temperatures close to $T_c$ and to rather massive gluons
in the partonic bulk matter. It will be of future
interest to perform a similar study at LHC energies for Pb+Pb reactions
since here a significantly larger $p_T$ range can be addressed and the
effect of gluon bremsstrahlung might become important again~\cite{Younus:2013rja}. Furthermore,
angular correlations between pairs of $D~{\bar D}$ mesons are expected to provide
further valuable information.

\section*{Acknowledgements}

The authors acknowledge inspiring discussions with J. Aichelin, P. B.
Gossiaux, C. M. Ko, O. Linnyk, R. Marty, V. Ozvenchuk, and R.
Vogt. This work was supported by DFG under contract BR 4000/3-1, and
by the LOEWE center "HIC for FAIR". JMTR is supported by the Program
TOGETHER from Region Pays de la Loire and the European I3-Hadron
Physics program. LT acknowledges support from the Ramon y Cajal
Research Programme and contracts FPA2010-16963 and FPA2013-43425-P
from Ministerio de Ciencia e Innovaci\'on, as well as from
FP7-PEOPLE-2011-CIG under Contract No. PCIG09-GA-2011-291679. The
computational resources have been provided by the LOEWE-CSC.

\end{document}